\documentclass{aa}
\usepackage{graphicx}
\graphicspath{{images/}}
\usepackage{txfonts}
\usepackage{booktabs}
\usepackage{orcidlink} 
\usepackage{hyperref}
\hypersetup{
    colorlinks=true,
    allcolors=blue,
    }

\newcommand{\kms}{\ensuremath{\rm km~s^{-1}}}

\newcommand{\msun}{\ensuremath{\rm M_{\odot}}}

\newcommand{\msunyr}{\ensuremath{\rm M_{\odot}~yr^{-1}}}

\newcommand{\mearth}{\ensuremath{\rm M_{\oplus}}}

\newcommand\xoutpars[1]{\let\helpcmd\xout\parhelp#1\par\relax\relax}
\newcommand\soutpars[1]{\let\helpcmd\sout\parhelp#1\par\relax\relax}

\begin{document}

   \title{Tidal phenomena in the Galactic Center: The curious case of X7}
   \author{
        Wasif~Shaqil\inst{\ref{inst1}} \fnmsep\thanks{Corresponding authors: \href{mailto:wasifshaqil@gmail.com}{wasifshaqil@gmail.com}}\orcidlink{0009-0009-4193-3724}
        \and
        Diego~Calderón\inst{\ref{inst2}}\fnmsep\inst{\ref{inst1}}\thanks{\href{mailto:calderon@mpa-garching.mpg.de}{calderon@mpa-garching.mpg.de}}\orcidlink{0000-0002-9019-9951}
        \and
        Stephan~Rosswog\inst{\ref{inst1}}\fnmsep\inst{\ref{inst3}}\orcidlink{0000-0002-3833-8520}
        \and
        Jorge~Cuadra\inst{\ref{inst4}}\fnmsep\inst{\ref{inst5}}\orcidlink{0000-0003-1965-3346}
        \and
        Anna~Ciurlo\inst{\ref{inst6}}\orcidlink{0000-0001-5800-3093}
        \and
        \\
        Mark~R.~Morris\inst{\ref{inst6}}\orcidlink{0000-0002-6753-2066}
        \and
        Randall~D.~Campbell\inst{\ref{inst7}}\orcidlink{0000-0002-3289-5203}
        \and
        Andrea~M.~Ghez\inst{\ref{inst6}}\orcidlink{0000-0003-3230-5055}
    }

   \institute{
        University of Hamburg, Hamburger Sternwarte, Gojenbergsweg 112, D-21029 Hamburg, Germany \label{inst1}
        \and
        Max-Planck-Institut für Astrophysik, Karl-Schwarzschild-Straße 1, 85748 Garching, Germany \label{inst2}
        \and
        The Oskar Klein Centre, Department of Astronomy, AlbaNova, Stockholm University, SE-106 91 Stockholm, Sweden \label{inst3}
        \and
        Departamento de Ciencias, Facultad de Artes Liberales, Universidad Adolfo Ibáñez, Av. Padre Hurtado 750, Viña del Mar, Chile \label{inst4}
        \and
        Millennium Nucleus on Transversal Research and Technology to Explore Supermassive Black Holes (TITANS) \label{inst5}
        \and
        Department of Physics and Astronomy, University of California, Los Angeles, CA 90095, USA \label{inst6}
        \and
        W. M. Keck Observatory, Waimea, HI 96743, USA \label{inst7}
    }

   \date{\today}

  \abstract
  {Several enigmatic dusty sources have been detected in the central parsec of the Galactic Center. 
  Among them is X7, located at only $\sim$0.02~pc from the central supermassive black hole, Sagittarius A* (Sgr A*).
  Recent observations have shown that X7 is becoming elongated due to the tidal forces of Sgr A*. 
  X7 is expected to be fully disrupted during its pericenter passage around 2035, which might impact the accretion rate of Sgr A*. 
  However, its origin and nature are still unknown.}
   {We investigated the tidal interaction of X7 with Sgr A* in order to constrain its origin. 
   We tested the hypothesis that X7 was produced by one of the observed stars with constrained dynamical properties in the vicinity of Sgr~A*.}
  {We employed a set of test-particle simulations to reproduce the observed structure and dynamics of X7. 
  The initial conditions of the models were obtained by extrapolating the observationally constrained orbits of X7 and the known stars into the past, making it possible to find the time and source of origin by minimizing the three-dimensional separation and velocity difference between them.}
   {Our results show that ejecta from the star S33/S0-30, launched in $\sim$1950, can to a large extent replicate the observed dynamics and structure of X7, provided that it is initially elongated with a velocity gradient across it, and with an initial maximum speed of $\sim$600~km~s$^{-1}$. 
   }
   {Our results show that a grazing collision between the star S33/S0-30 and a field object such as a stellar-mass black hole or a Jupiter-mass object is a viable scenario to explain the origin of X7. 
   Despite the uncertainties in the rate of these encounters, recent estimations show that it is plausible for such a scenario to have occurred recently.}

\authorrunning{Wasif Shaqil}
   \maketitle

\section{Introduction}

    The Milky Way Galactic Center (GC) harbors the closest supermassive black hole (SMBH) to Earth: Sagittarius~A* (Sgr~A*), which has a mass of $4.3 \times 10^6$ $\msun$ at a distance of 8.3~kpc \citep{GRAVITY_2022}. 
    This makes it a unique laboratory to study the detailed orbital motion of the stellar and gaseous components in the vicinity of a SMBH \citep[see][for a review]{genzel_2010,ciurlo2025}. 
    Over 30 years of near-infrared (NIR) observations have enabled the monitoring of hundreds of stars with high precision \citep{2002_Schoedel, 2003_Ghez, 2008_Ghez, 2009_gillessen, Gillessen_2017, vonFellenberg_2022}, and even the successful testing of the predictions of general relativity \citep[e.g.,][]{Do_2019, 2019_Gravity, 2020_Gravity}. 
    Currently, there are 195 stars with constrained orbits within the central parsec \citep{vonFellenberg_2022}. 
    Many of them are B-type stars \citep{Eisenhauer_2005}; there are also tens of O-type and Wolf-Rayet (WR) stars that have significant mass loss in the form of winds \citep{paumard2006, Martins2007, Habibi_2019}. 
    Over the last decade, a lot of attention has been paid to the dusty sources known as G objects that coexist with the stars in the vicinity of Sgr~A* \citep[e.g.,][]{ciurlo2023}. 
    Despite long observation campaigns and theoretical efforts, their nature remains unknown \cite[see][for a review]{mapelli2016, 2020_ciurlo}.

\begin{figure*}
    \centering
    \includegraphics[width=0.9\linewidth]{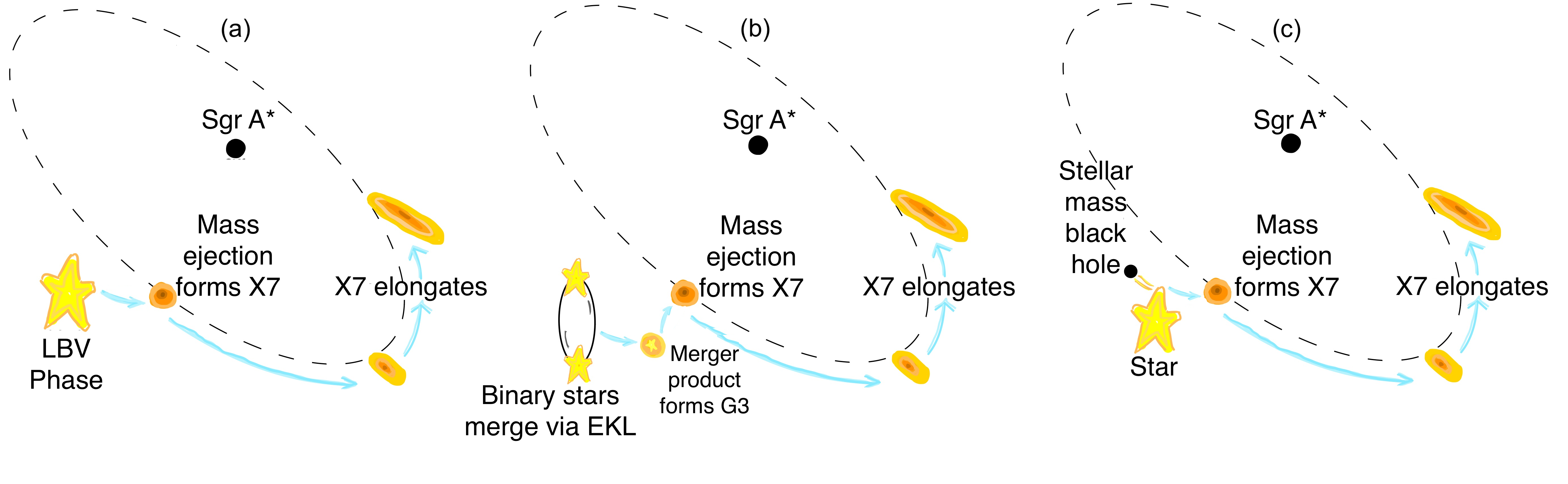}
    \caption{Schematic representation of the three hypotheses for the origin of X7 studied in this work. (a) X7 formed from a stellar wind of a star during an episode of high mass loss such as a LBV phase. (b) X7 as the ejecta from a stellar merger via the EKL mechanism, where G3 is the merger product due to similar orbital motion. (c) X7 as the ejecta from a collision of a star with a field object such as a stellar-mass black hole or a Jupiter-mass object. Bear in mind that the representations are not to scale.}
    \label{fig:schematic}
\end{figure*}

    \cite{Gillessen_2012} discovered the gaseous and dusty source G2, moving on a highly eccentric orbit toward Sgr~A*.
    Initially, the source was interpreted as a purely gaseous object with a total mass of merely 3M$_{\oplus}$. 
    Its observed tidal interaction over timescales on the order of years attracted attention due to its potential effect on the quiescent accretion state of Sgr~A*. 
    However, there was no clear observed enhancement in the activity of Sgr~A* in X-rays \citep{bouffard2019}, but a bright flare in the NIR in 2019 could potentially be attributed to it \citep{do2019b, Paugnat_2024}. 
    Over the years, many objects with similar characteristics have been observed in the central 0.1~pc, increasing the known population of G objects to ten \citep[e.g.,][]{pfuhl2015,witzel2017,2020_ciurlo,peissker2019, Peissker2024}. 
    Many show extended emission of the Brackett gamma (Br$\gamma$) recombination line at 2.1661~$\mu$m and point-source emission in the L$'$ band at 3.776~$\mu$m. 
    None show obvious hints of a stellar object within the extended source. 
    As a result, theoretical efforts have focused on constraining their true nature, especially that of G2, but no consensus has been reached on its nature and origin.
    The main debate is focused on whether G2 is a purely gaseous cloud or harbors a compact (stellar) object. 
    Regarding the purely gaseous cloud hypothesis, it has been suggested that G2 could be a gas clump formed in stellar wind collisions \citep{burkert_2012,Calderon_2016}, the result of a slow wind from a luminous blue variable (LBV) star \citep{burkert_2012}, or a nova outburst from the partial tidal disruption of a giant star \citep{meyer2012}, among others. 
    Within the compact object argument, G2 has been hypothesized to be an evaporating circumstellar disk \citep{miralda2012,murray2012}, a protoplanet \citep{mapelli2015}, a mass-losing low-mass star \citep{ballone2013,decolle2014,2015_valencia,ballone2016,ballone2018}, the product of a stellar merger \citep{2014_witzel,2015_Prodan, 2020_ciurlo}, or a young stellar object \citep[][]{Peissker2024}. 
    Moreover, \cite{Peissker_2024b} also argue that there are binaries within the G object population. 

    In this region, at a separation of $\sim$0.5\arcsec ($\sim$0.02~pc) from Sgr~A*, the source X7 has been observed to be interacting tidally with the central black hole. 
    Although X7 has certain similarities with the G objects, for example it is an extended source composed of gas and dust, it is much larger, reaching  $\sim$3300~au in 2021, whereas G objects are on the order of $\sim$100~au. 
    It was reported for the first time by \cite{clenet2004} and since then has been routinely monitored by several groups \citep{muzic2007,peissker2019, peissker2021,ciurlo2023}. 
    Initially, it was thought to be a bow shock, but recent observations suggest a morphological deviation from that picture \citep{ciurlo2023}. 
    \cite{peissker2021} propose that X7 is a circumstellar envelope of the star S50/S0-73 since they were close in earlier observations. 
    However, \cite{ciurlo2023} argue against that hypothesis; they find that there is a significant three-dimensional spatial and dynamical separation between them. 
    Additionally, they estimate that the pericenter passage of X7 will take place around 2035. 
    If X7 is a purely gaseous and dusty source, we will be able to witness the tidal disruption of this object by Sgr~A* and, consequently, the enhanced accretion activity of Sgr A*.  
    An alternative scenario has recently been proposed by \cite{Peissker2024}, who suggest that both X7 and a similar dusty feature called X3 are young stellar objects. 
    Nevertheless, the nature of X7 is still unknown, and observational and theoretical efforts are ongoing to unravel its mystery.

    We have carried out a study to constrain the origin of X7. 
    We worked under the assumption that the source is purely gaseous and dusty and, therefore, that its age is less than its orbital period of 200~years. 
    We present a set of test-particle simulations and analytical estimates to study how likely it is that X7 was related to any of the stars in the central parsec of the GC. 
    Our results show that the current observations of X7 are consistent with the source being the ejecta resulting from a grazing collision between the S star S33/S0-30 (an early B-type star) and a field object such as a stellar-mass black hole or a Jupiter-mass object. 
    It has been suggested that X7 could be connected to the G object G3 \citep{ciurlo2023} because of the remarkable alignment of their orbits. 
    However, we find that they are most likely not connected since, 
    provided the age of X7 is less than 200~yrs, they show significantly different orbital phases, even when considering the large associated uncertainties. 
    ?

This article is organized as follows: In Sect.~\ref{X7} we present the properties of X7 as well as the three hypotheses for its formation studied in this work. 
    Section~\ref{methods} contains the details of the calculations, the methods, and the numerical setup for testing the hypotheses. 
    In Sect.~\ref{results} we present the results from our analysis and simulations. 
    Section~\ref{discuss} discusses the implication of our results for understanding the origin of X7. 
    Finally, we summarize and conclude this study in  Sect.~\ref{conclusion}. 
    Throughout this paper, we use $M_\text{BH}=4.3\times 10^6~\text{M$_{\odot}$}$ and $ R_0 = 8.3~\text{kpc}$, where $M_\text{BH}$ and $R_0$ are the mass of Sgr~A* and the distance to it from Earth \citep{GRAVITY_2022}. 

            \begin{figure*}
                \centering
                \includegraphics[width=0.85\linewidth]{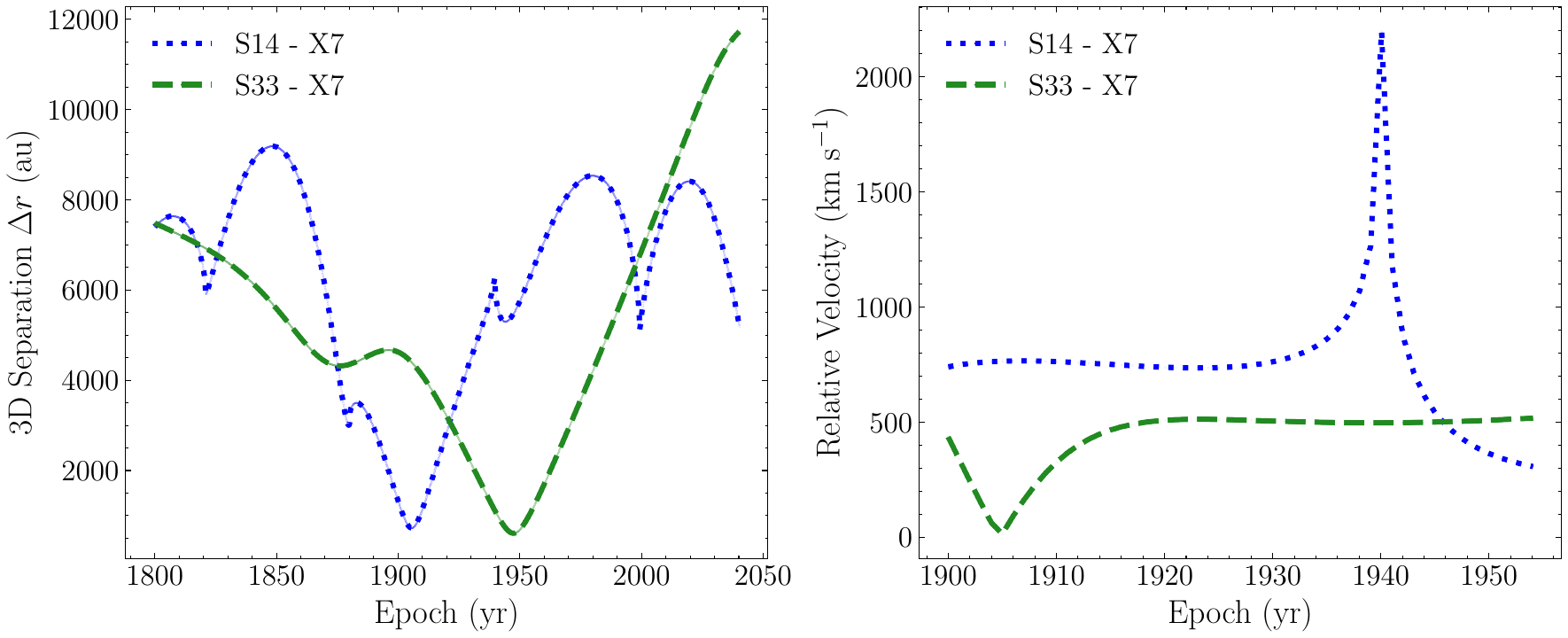}
                \caption{
                Three-dimensional separation (left) and relative velocity (right) between stars and X7 as a function of time. 
                The two stars closest to X7, S14/S0-16 and S33/S0-30, are shown. 
                 S14/S0-16-X7 is displayed as a dotted blue line and S33/S0-30-X7 as a dashed green line. 
                The calculations are shown in the time periods 1800-2050 (left) and 1900-1950 (right).}
                \label{fig:radsep_relvel}
            \end{figure*}

    \section{Hypotheses for the origin of X7}
    \label{X7}

        X7 has been monitored continuously since 2002 with telescopes at the W.M. Keck Observatory \citep{ciurlo2023} as part of the Galactic Center Orbits Initiative (GCOI; PI: Andrea Ghez). 
        The data traced the dust thermal emission – through images in the L$^\prime$ band (3.776 $\mu$m) obtained with the imager NIRC2 -- and the gas emission through the Br$\gamma$ hydrogen recombination line observed with the integral field spectrograph OSIRIS \citep{larkin2006}. 
        These observations have shown that the size of X7 has nearly doubled from $L\sim 2000$~au in 2003 to $L\sim 3300$~au in 2021.
        Additionally, it was possible to fit a Keplerian orbit to its forward tip, revealing a semimajor axis of $4800\pm1100$~au and an eccentricity of $0.34\pm0.05$ \citep{ciurlo2023}.
        Under the assumption that it is a purely gas and dust feature, a total mass of $\sim$50~M$_{\oplus}$ was derived using the observed Br$\gamma$ flux. 
        Then, it is straightforward to estimate that the ratio between the derived and Roche densities (the critical density to remain intact against tidal forces of the SMBH) of X7 is very small ($\sim$10$^{-5}$). 
        This shows that its self-gravity is negligible provided that no compact source is associated with it. 
        Based on this, its age is constrained to be less than its orbital period ($\sim$200~yr), as otherwise the tidal forces would have already destroyed it. 

        Under the assumption that X7 is only made out of gas and dust, we propose and study three hypotheses to explain its origin. 
        \begin{itemize}
            \item First, X7 could be the result of a mass outflow from a massive star going through a major mass-loss episode such as a LBV phase. 
            This scenario is plausible since the central parsec hosts hundreds of O- and B-type stars \citep[e.g.,][]{paumard2006,Martins2007, Habibi_2019}.

            \item The second hypothesis conceives X7 as ejecta from a binary merger process.  
            This case is supported by the fact that the sources G3 and X7 could be considered dynamically linked as they move on very similar orbits and share common observational properties \citep[see][]{2020_ciurlo, ciurlo2023}. 
            In this scenario, G3 would be the merger product and X7 the mass ejecta from the eccentric Kozai-Lidov (EKL)-induced merger process \citep[][]{kozai1962,lidov1962, Naoz2016}.

            \item Third, X7 could be ejecta from a grazing collision of a star with a field object such as a stellar-mass black hole or a Jupiter-mass object \citep{ciurlo2023}. 
            Here, X7 would be the stripped material unbound from the star during the collision. 
            This is plausible as the stars in the GC are expected to undergo multiple stellar collisions due to the high stellar density \citep{1996_genzel, 2009_Dale}. 
            Moreover, the central parsec should contain a large population of stellar-mass black holes due to dynamical mass segregation \citep{1993_Morris, 2006_Freitag}.
        \end{itemize}
        Schematic representations of the hypotheses are shown in Fig.~\ref{fig:schematic}.

            \begin{figure*}
            \sidecaption
                \centering
                \includegraphics[width=0.7\linewidth]{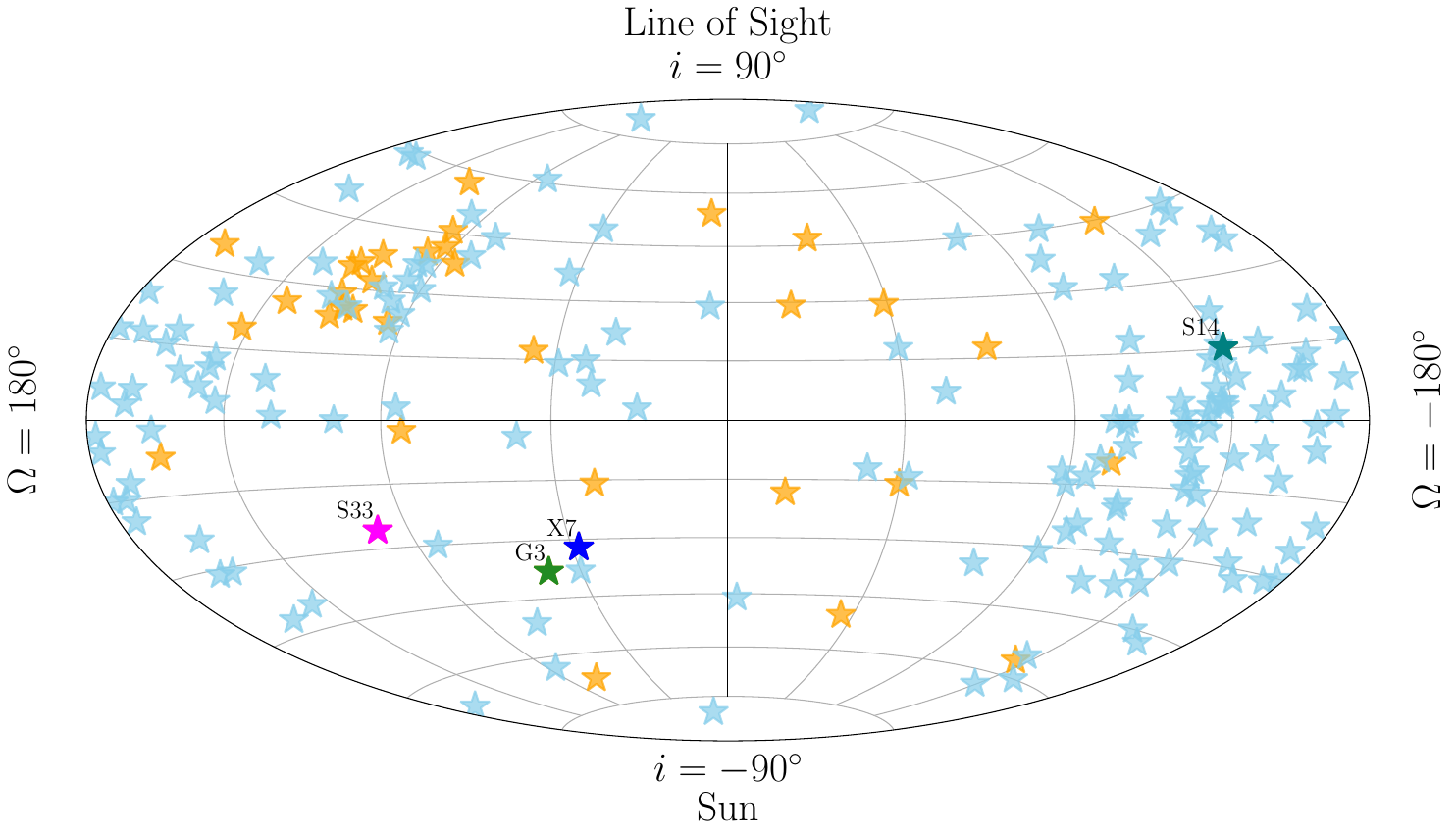}
                \caption{
                Hammer projection of the orientation of the angular momentum vectors of the stars with full  (orange stars) and incomplete (sky-blue stars) orbital solutions. 
                The vertical dimension represents the inclination of the orbit ($i$), and the horizontal dimension represents the longitude of the ascending node ($\Omega$). 
                An edge-on star in a clockwise orbit will be at the top of the hammer projection, and a face-on star will be at the equator.
                The bottom half shows the counterclockwise orbits. 
                S14/S0-16, S33/S0-30, X7, and G3 are shown as teal, magenta, blue, and green stars, respectively.
                }
                \label{fig:Angular Momentum}
            \end{figure*}

    \section{Analysis}
    \label{methods}

        Regardless of the hypothesis considered, all of them have in common that X7 would be related to a stellar source in the region. 
        Hence, as a first step we investigated which stars could be linked to X7 based on their dynamics. 
        It is to be noted that when we mention the orbit of X7, we are referring to the orbit of its tip \citep{ciurlo2023}. 
        Once found find the best star candidate to be related to X7, we simulated the dynamical evolution of a cloud of test particles launched from the position and time of the closest encounter. 
        In this section we describe in detail these calculations.
    
        \subsection{Correlation with stars in the Galactic Center}\label{MINECC}

            First, we analyzed whether X7 is related to one of the stars in the vicinity of Sgr A*. 
            Thus, we searched for stars that were close simultaneously in space and time from X7 by tracing their orbits back in time into the past. 
            To do so, we assumed that the orbits have not changed significantly in the last 200 years. 
            This is a reasonable assumption since the dominant gravitational field is due to Sgr A*.
            Currently, there are 36 stars with fully constrained orbital parameters and 159 stars with constrained orbits but incomplete orbital solutions ($z$-coordinate missing; \citealt{vonFellenberg_2022}). 
            Following the \texttt{MINECC} method in \cite{cuadra2008}, we looked for the minimum eccentricity possible for each of the 159 stars and considered the corresponding value of $z$ in order to model their orbits. 
            By doing so, we have a total of 195 stars with orbits for our analysis.
             
            We traced back their orbits for the period $1800-2024$ and selected the stars having the closest three-dimensional distance to X7 at some point during this time.
            To select the best candidates, we chose that the three-dimensional distance of closest approach must be $<$1000~au, and their relative velocities at that time had to be $<$1000~km~s$^{-1}$. 
            Following this procedure, we found two stars: S14/S0-16 and S33/S0-30, both of which have fully determined orbital solutions \citep{vonFellenberg_2022}.
            Figure \ref{fig:radsep_relvel} shows the three-dimensional separation (left panel) and relative velocity (right panel) of these two stars with respect to X7 as a function of time during the period $1800-2050$.
            In the case of S14/S0-16 (dotted blue line), the closest approach to X7 was in $\sim$1905 at a separation and a relative velocity of $\Delta r$ $\sim$727~au and $\Delta v$ $\sim$760~km~s$^{-1}$, respectively. 
            In the case of S33/S0-30 (dashed green line), the shortest distance was $\Delta r\sim610$~au and the relative velocity $\Delta v\sim505$~km~s$^{-1}$ in the year $\sim$1947. 
            It is to be noted that other assumptions to constrain the missing $z$ coordinate of the stars with incomplete orbits were explored. 
            Specifically, we sampled all possible values of $z$ compatible with the observational constraints and found no other candidate that satisfied the criteria.
            
            As a next step, we inspected the angular momentum direction of the candidate stars and X7. 
            Figure {\ref{fig:Angular Momentum}} is a Hammer projection that shows the orbital angular momentum direction for all stars in our sample as well as the sources X7 and G3 \citep{2020_ciurlo, ciurlo2023}. 
            Notice that S14/S0-16 has a very different angular momentum orientation compared to both X7 and G3, whereas S33/S0-30 is much closer to X7 and G3 in the Hammer projection. 
            In principle, this indicates that S33/S0-30 is a more promising candidate for being related to X7.
            Despite some stars having a similar angular momentum orientation to that of X7, it is important to bear in mind that they have a large minimum three-dimensional separation to X7 ($>6000$~au), so they are unlikely to be related. 
            \cite{2009_gillessen, Gillessen_2017} have characterized the star S33/S0-30 as an early-type star that is part of the S-star cluster. 
            Unfortunately, we are not aware of additional information on this star.

            \begin{figure}
                \centering
                \includegraphics[width=0.85\linewidth]{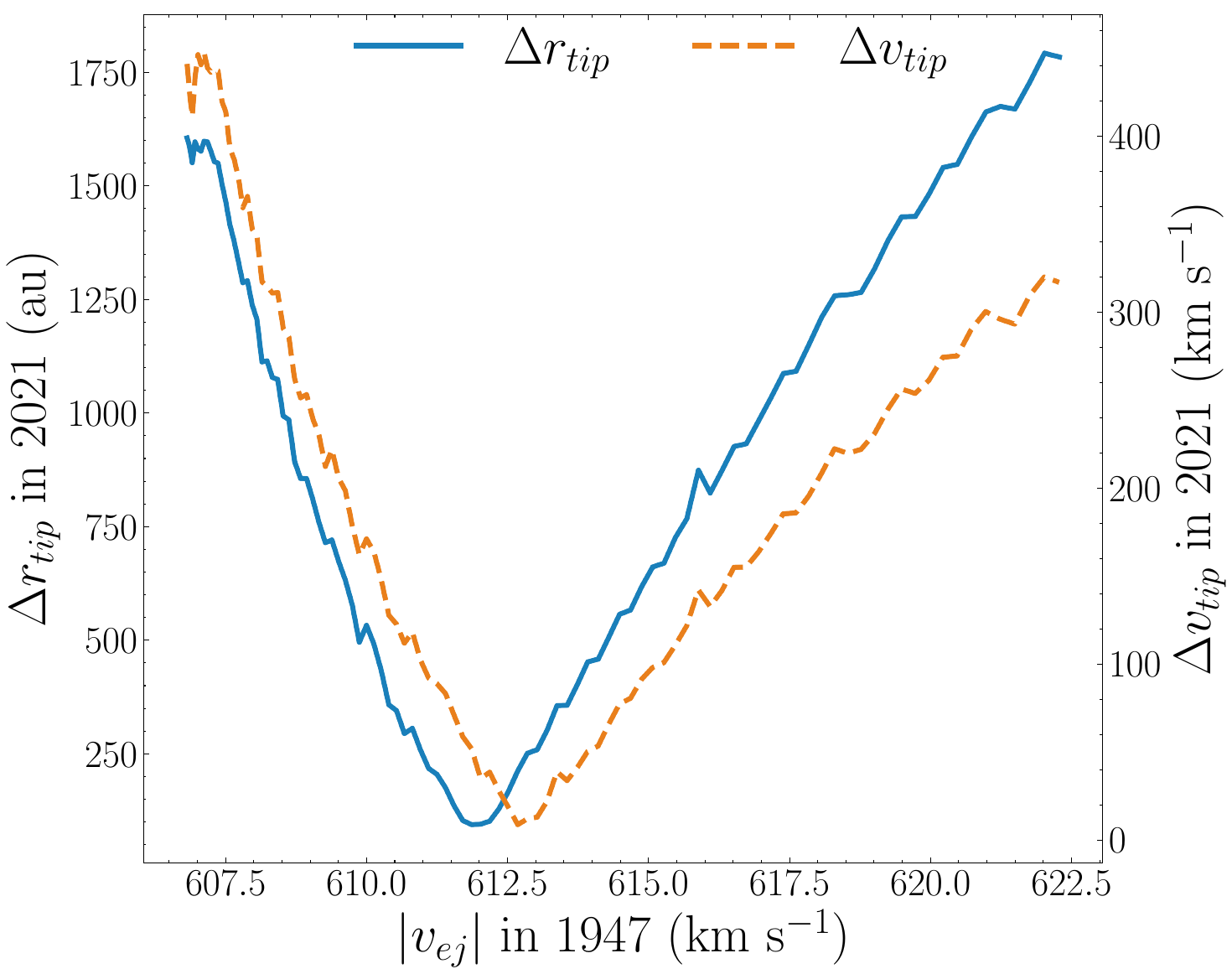}
                \caption{
                Three-dimensional separation ($\Delta r_{\text{tip}}$) and relative velocity ($\Delta v_{\text{tip}}$) between the tip of the simulated cloud and X7 in 2021 as a function of the magnitude of initial velocity ($|v_\text{ej}|$) of the simulated cloud.
                The solid blue and dashed orange lines represent $\Delta r_\text{tip}$ and $\Delta v_\text{tip}$, respectively.
                Notice that both quantities have a minimum at $|v_\text{ej}|\sim612~\kms$. 
                }
                \label{fig:v_vs_sep_vs_relvel} 
            \end{figure}

        \subsection{Test-particle simulations}
        
            We modeled the evolution of a cloud represented with test particles launched from the position and time of the closest encounter between S33/S0-30 and X7. 
            Here we describe the numerical setup and technique. 
            Then, we show how we calculated the initial conditions for the simulations, such as the initial velocity, direction, radius, and shape of the cloud.

             \begin{figure}
                \centering
                \includegraphics[width=0.85\linewidth]{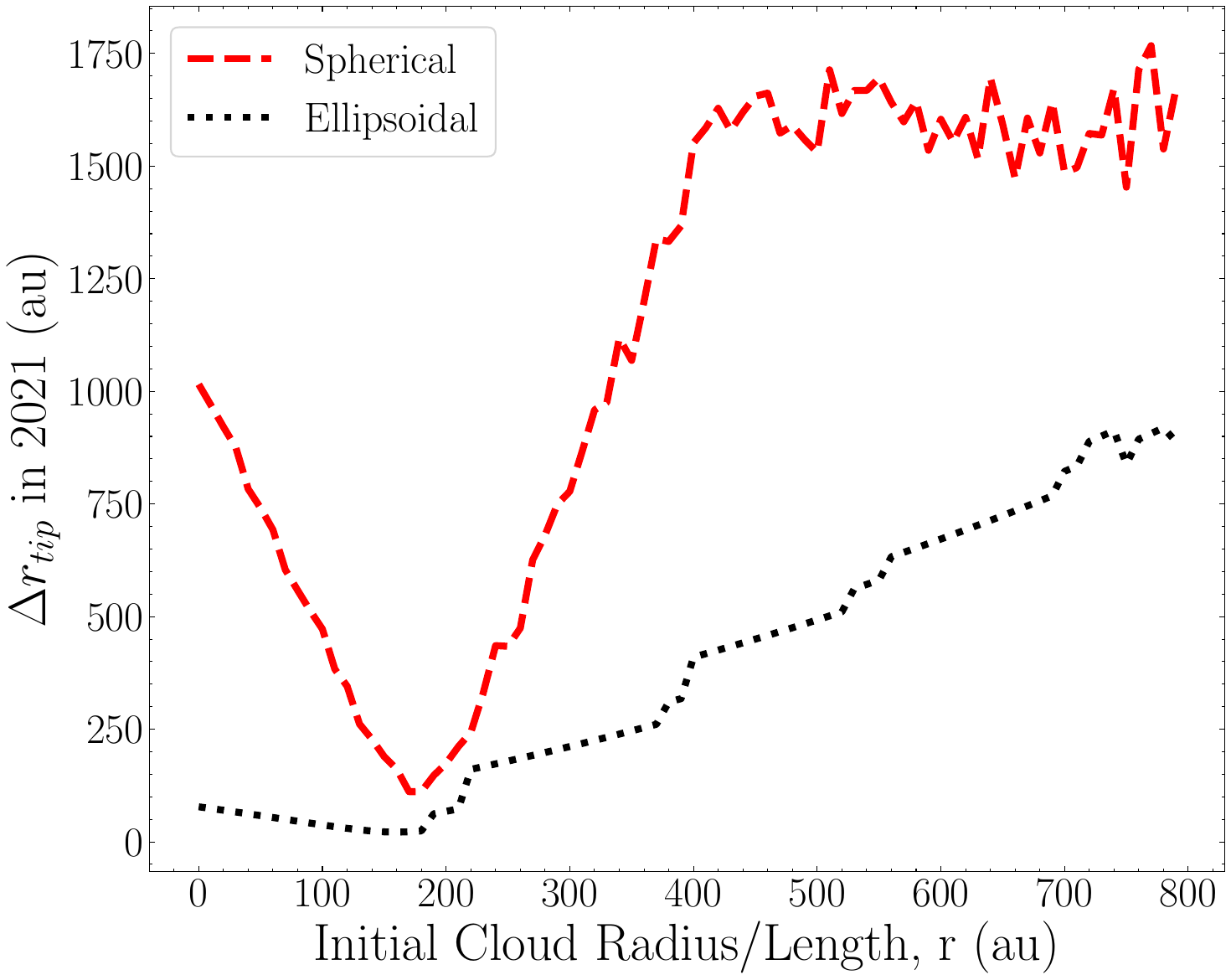}
                \caption{Three-dimensional separation of the tip of the simulated cloud and X7 in 2021 ($\Delta r_{\text{tip}}$) as a function of the initial size of the simulated cloud. The dotted black and dashed red lines represent the initial length of the ellipsoidal and initial radius of the spherical clouds, respectively.}
                \label{fig:initial_R}
            \end{figure}

            \begin{table}
                \centering
                \caption{Simulation results using different initial configurations.}
                \begin{tabular}{llllll}
                    \toprule
                    Sim ID & $N_{\text{runs}}$ & $N_{\text{particles}}$ & $\Delta r_{\text{tip}}$ & $\Delta v_{\text{tip}}$ \\
                    & & & (au) & (km~s$^{-1}$) \\
                    \midrule
                    sp$_{\text{uniform}}$ & 10000 & 10-10000 & 116$\pm$8 & 42$\pm$6 \\
                    sp$_{\text{Gaussian}}$ & 10000 & 10-10000 & 250$\pm$21 & 102$\pm$6 \\
                    Ellipsoidal & 10000 & 10-10000 & 50$\pm$21 & 22$\pm$6\\
                    \bottomrule
                \end{tabular}
                \label{tab:SIMRUNS}
                \tablefoot{
                All simulations are starting from the initial position of S33/S0-30 in 1947, and running till the present time. The uncertainties are 1$\sigma$ deviations from the mean based on 10000 simulation runs.
                }
            \end{table}
        
        \label{Sim_setup}
        \subsubsection{Numerical setup}

        The simulations considered the dynamical evolution of a cloud represented with test particles under the influence of the gravitational field of Sgr~A*. 
        The equations of motion were integrated numerically using a leapfrog algorithm with a timestep that is a small fraction of the circular orbital timescale at separation $R$, 
            \[    
                \Delta t = C \sqrt{\frac{R^3}{GM_\text{BH}},}
            \]
        where $R$ is the radial distance of the particle to the location of Sgr~A*, $M_\text{BH}$ is the mass of the SMBH, $G$ is the gravitational constant, and $C$ is the prefactor, which is set to $C=0.01$ for accuracy. 
        We applied an adaptive time-stepping scheme in which the minimum timestep, $\Delta t$, of all the particles at each iteration is considered. 
        The initial position of the cloud was set to the position of S33/S0-30 in 1947. 
        Then, it is only necessary to specify the size of the cloud and its initial velocity.

        \subsubsection{Initial velocity}
        \label{select}
            To find the initial velocity of the cloud, we calculated the velocity required for ejecta from the position ($\vec{r}$) of S33/S0-30 in 1947 to achieve the same observed specific angular momentum ($\Vec{l}$) of the tip of X7. 
            To do so, we solved the following equation to find $\vec{v}_\text{X7}(t=1947)$,
            \begin{equation}
                \vec{l}_\text{X7} = \vec{r}_\text{S33/S0-30}(t=1947) \times \vec{v}_\text{X7}(t=1947)
            .\end{equation}
            This corresponds to a system of linear equations obtained that is not independent, so a unique solution is not possible. 
            However, for each value of $v_z$ there are unique values of $v_x$ and $v_y$. 
            The components $v_x$ and $v_y$ can be obtained from the following equations:
            \begin{eqnarray}
                \centering
                v_x
                &=& -\left[442 + \left(\frac{v_z}{1.8~\text{km}~\text{s}^{-1}}\right)\right] ~\text{km}~\text{s}^{-1},
                \\
                v_y
                &=& \left[699 + \left(\frac{v_x}{1.6~\text{km}~\text{s}^{-1}}\right)\right]~\text{km}~\text{s}^{-1}.
            \end{eqnarray}
            Then, we sampled a wide range of values for $v_z$ from $-$10000~km~s$^{-1}$ and 10000~km~s$^{-1}$ with a step size of 1~$\kms$. 
            In this way, the velocity magnitude will be of the same order as the typical velocities of the stars in the GC.           
            To select the optimal initial velocity and its direction, we selected the velocities that minimize the three-dimensional separation and relative three-dimensional velocity between the tip of the simulated cloud and the observed tip of X7 in 2021, which are represented as $\Delta r_{\text{tip}}$ and $\Delta v_\text{tip}$, respectively. 
            Figure \ref{fig:v_vs_sep_vs_relvel} shows $\Delta r_{\text{tip}}$ and $\Delta v_\text{tip}$ as a function of the magnitude of the initial velocity ($|\vec{v}_\text{ej}|$) in 1947. 
            It is important to remark that only a zoomed version of the region around the minima is shown. 
            Here it is possible to observe that an initial velocity of $\sim612~\kms$ minimizes both the $\Delta r_{\text{tip}}$ and $\Delta v_\text{tip}$.
            
            For a range of $v_z$,  $(-40 < v_z < -30)\ \kms$, the optimal range of components of initial velocity, having an initial velocity magnitude of $\sim 610\ \kms$, is between
            \begin{equation}
                (-426 < v_x < -420)\  \kms \quad and \quad  (433 < v_y < 436) \ \kms
            .\end{equation}
            By "optimal," we mean that the separation of the tip of the simulated cloud and the observed tip of X7 ($\Delta r_{\text{tip}}$) is less than 175 au ($\sim$21 mas), which falls within the margin of error in the 2021 observations from \cite{ciurlo2023}. 

        \subsubsection{Cloud structure and size}

            We investigated the cases of a spherical cloud uniformly sampled, a spherical cloud sampled following a three-dimensional Gaussian distribution, and an ellipsoidal cloud uniformly sampled.
            Here we focus on the case of the ellipsoidal cloud as it gave the most relevant results. 
            We refer the reader to Appendix~\ref{spherical} to see the results of the spherical cloud cases. 
            In the ellipsoidal cloud case, we set up a cloud with $N$ particles that are uniformly distributed along a length of $L = 200$ au, with the semimajor axis $a=L/2$. 
            There is negligible dependence on the other two semimajor axes ($b$ and $c$), as long as they are less than $a$. 
            Moreover, for each particle, we set the polar angle $0^\circ < \theta < 90^\circ$ and the azimuthal angle $0^\circ < \phi < 180^\circ$. These two angles have, in practice, no effect on our overall results. 
            The initial position of the tip of the cloud $\vec{r}_\text{tip}^\text{initial} = (x_0,\ y_0,\ z_0)$ is assigned from the position of S33/S0-30 in 1947. 
            The position of the rest of the $N-1$ particles were assigned by adding the position of each particle from the center of the ellipsoid ($x, y, z$) to the initial position, 
            \[
                \texttt{initial\_position[i]} = (x_0 + x[i], y_0 + y[i], z_0 + z[i]).
            \]

            To assign the initial velocity of each particle, we used a coefficient of linear increase ($k$) in au yr$^{-1}$ dx$^{-1}$ such that there is a velocity spread of $\sim$100~$\kms$ from tip to tail. 
            Moreover, we used an initial position angle of $\theta_\text{ini} = 75^\circ$ for the ridge of the ellipsoid obtained after testing angles in the range $0^\circ < \theta <90^\circ$.   
            To find the initial cloud size, we tested initial lengths from 1 au to 800 au and analyzed the three dimensional separation of the tip of the simulated cloud and X7 in 2021.  
            Figure \ref{fig:initial_R} shows $\Delta r_{\text{tip}}$ as a function of the initial ellipsoid length for the ellipsoidal and spherical cases. 
            In both the cases, it is possible to observe that an initial length/radius of $\sim$200~au minimizes $\Delta r_\text{tip}$ in 2021.

            In Table \ref{tab:SIMRUNS}, we have summarized the different initial configurations for our simulations, their corresponding three-dimensional separation, and relative velocity in 2021. 
            We also tested different number of particles to represent the cloud ($N_\text{particles}$), between 10 and 10000. 
            We find that this does not affect the results in a significant way.

\section{Results}
\label{results}

    \begin{figure}
        \centering
        \includegraphics[width=0.85\linewidth]{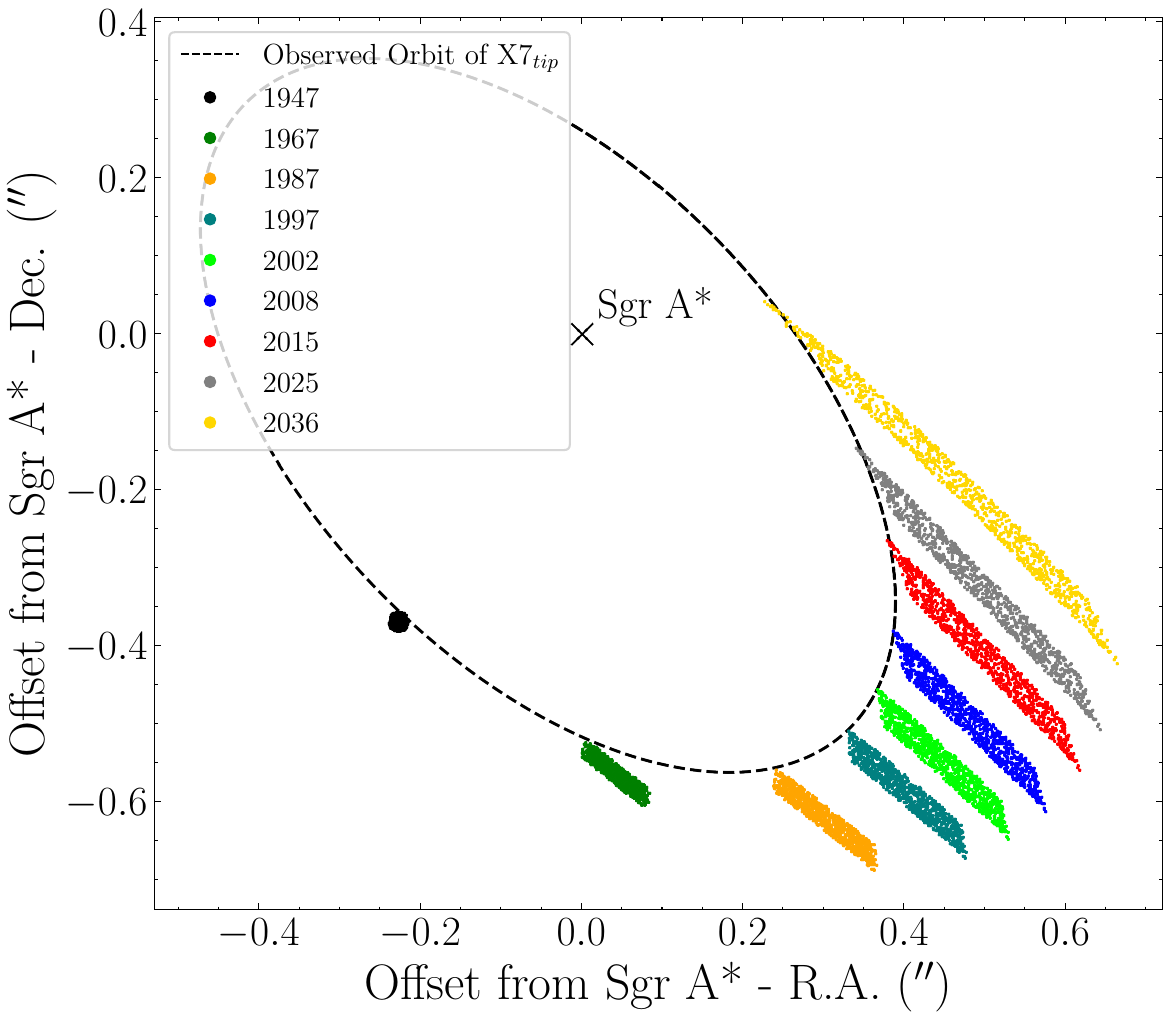}
        \caption{
        Test-particle simulation of a uniformly sampled ellipsoidal cloud launched from the position of S33/S0-30 in 1947. 
        Each color represents different times in its dynamical evolution.
        The sky-projected observed orbit of the tip of X7 is represented as the dashed black line. 
        As a reference, Sgr~A* is marked with a cross at the origin.}
        \label{fig:sim_cloud}
    \end{figure}

    In this section we present the results of the simulations with the optimal initial conditions obtained previously. 
    The result of one of the realizations is presented in Fig.~\ref{fig:sim_cloud}, which shows the sky-projected evolution of the initially ellipsoidal cloud at different simulation times. 
    We find that the cloud becomes more elongated through time and maintains a roughly constant position angle of $\sim$45$^{\circ}$ from 2002 until its pericenter passage around $\sim$2035.

        \begin{figure}
            \centering
            \includegraphics[width=0.85\linewidth]{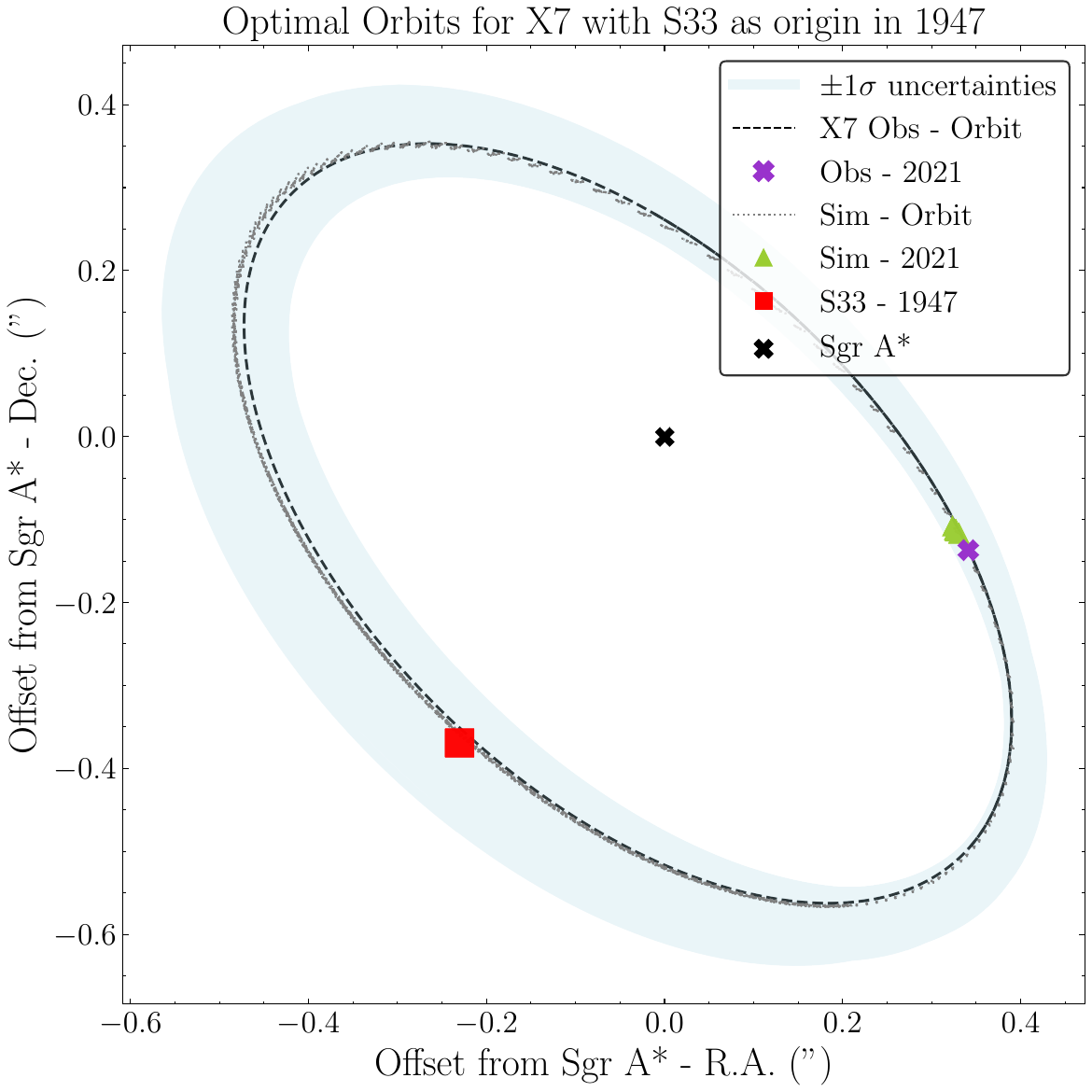}
            \caption{
            Comparison of the sky-projected orbits and position in 2021 of X7 and the tip of the simulated clouds.
            The dashed black line shows the best-fit orbit of the tip of X7, the light blue shaded region shows the uncertainties with a 68$\%$ confidence interval, and the purple cross marks its position in 2021 \citep{ciurlo2023}. 
            The dotted gray lines show the orbit of the simulated clouds for each of the optimal velocities. 
            The red squares and light green triangles show the position of the tip of the simulated clouds in 1947 and 2021, respectively. 
            }
            \label{fig:best-cases-X7}
        \end{figure}

        \begin{figure*}
            \centering
            \includegraphics[width=0.75\linewidth]{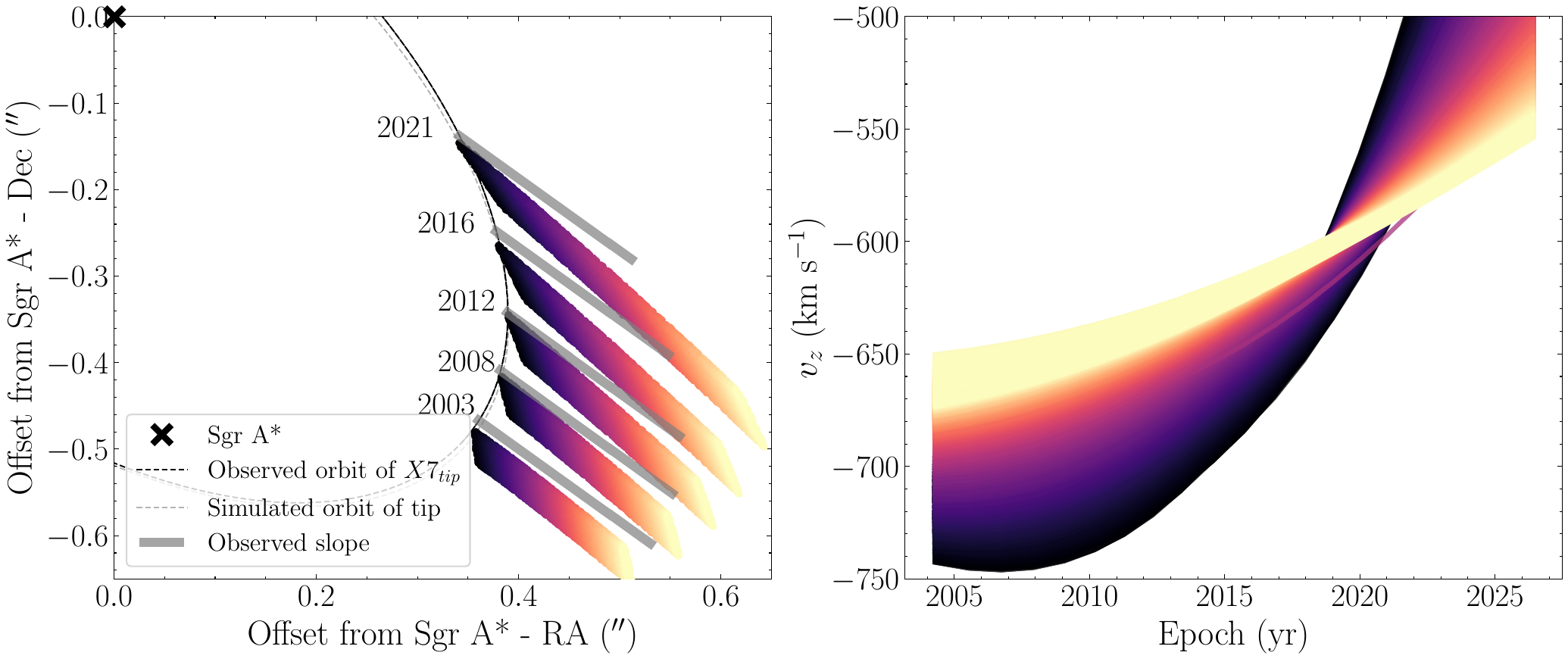}
            \caption{
            Evolution of the inclination and line-of-sight velocity for the best-fitting simulated cloud during the period of the observations.
            Left panel: Sky projection of the simulated cloud at $t=2003$, 2008, 2012, 2016, and 2021~yr. 
            The particles are colored from tip to tail from black to yellow.
            The gray shaded lines show the observed slope of the ridge of X7 at the same epochs (length not to scale). 
            Right panel: Line-of-sight velocity of the particles as a function of time.
            The colors represent the corresponding particles shown in the left panel. 
            This figure is analogous to Fig. 10 of \cite{ciurlo2023}.}
        \label{fig:ellipse}
        \end{figure*}

    \subsection{X7 modeled evolution with S33/S0-30 as the progenitor}
    \label{S33/S0-30-X7}
        
        Figure \ref{fig:best-cases-X7} shows the sky-projected orbits of X7 and the tip of the simulated cloud for each set of optimal velocities. 
        The sky-projected orbits have a strong correspondence (by construction) with the observed orbit, the three-dimensional spatial difference of $\sim$50~au (see Table \ref{tab:SIMRUNS}), is well within the uncertainty as reported by \cite{ciurlo2023}. 
        Moreover, the best realization implies a position of the tip of the simulated cloud at 0.348$\arcsec$ and -0.151$\arcsec$ in 2021 in offsets from Sgr~A* in right ascension and declination, respectively, both of which are within the error bars of the 2021 measurements by \cite{ciurlo2023}.

        Figure \ref{fig:ellipse} shows the spatial and dynamical evolution of the simulated cloud during the period of the observations from 2002 to 2021. 
        The left panel shows the sky-projected morphology evolution of the cloud. 
        This configuration is obtained only if the ellipsoidal cloud is initialized with a initial length $\sim$200~au, and position angle $\theta_\text{ini} \sim 75^\circ$. 
        Notice that the simulated cloud maintains a relatively constant orientation throughout the observational period from 2002 to 2021. 
        Although the agreement with the observed orientation of X7 is not perfect, it is off by $< 10^{\circ}$. 
        Additionally, the length of the simulated cloud is consistent with the observed length of X7 in 2021, which is $\sim$3300~au. 
        The right panel shows the line-of-sight velocity of the simulated cloud as a function of time. 
        This analysis shows that there is a crossover of the radial velocity from tip to tail around 2018. 
        In this case, the radial velocity of the tip decelerates by $\sim200 \ \kms$ (see the evolution of the black band), whereas the relative velocity of the tail remains relatively constant till $\sim$2021 (see the evolution of the yellow markers). 
        It is important to remark that the observations of X7 shows exactly the same behavior although at a slightly different time \citep{ciurlo2023}.

        \begin{figure*}
            \centering
            \includegraphics[width=0.75\linewidth]{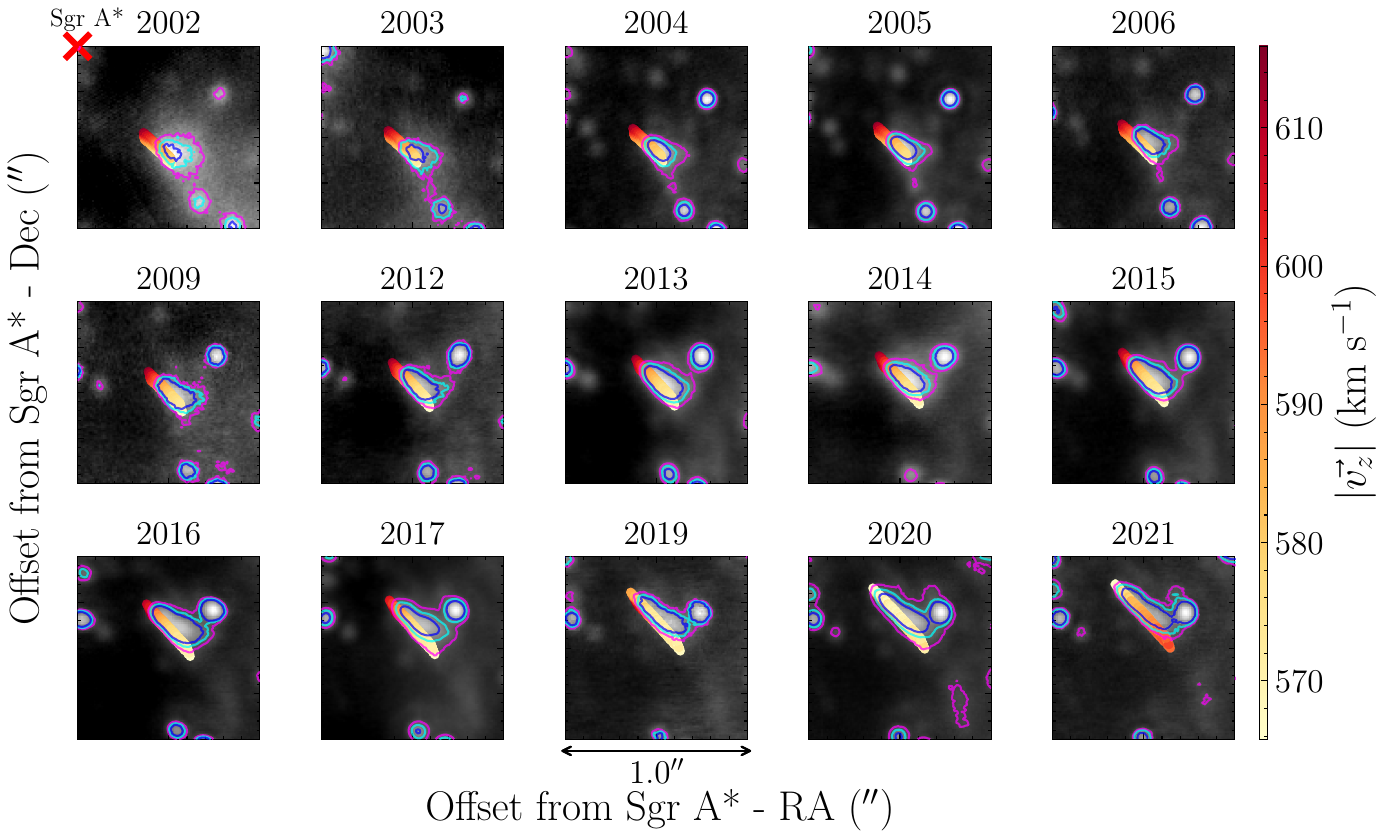}
            \caption{Comparison of the morphological evolution of X7 and the simulated cloud. 
            The panels show 1$\arcsec\times$~1$\arcsec$ L$'$-band images highlighting the emission from X7 with contours in the period 2002-2021.
            The simulated cloud is overlaid with colored markers that encode the line-of-sight velocity. 
            The position of Sgr A* is on the top left corner of each panel. 
            North is up and east to the left.
            }
            \label{fig:angled_combi}
        \end{figure*}

        \begin{figure}
            \centering
            \includegraphics[width=0.85\linewidth]{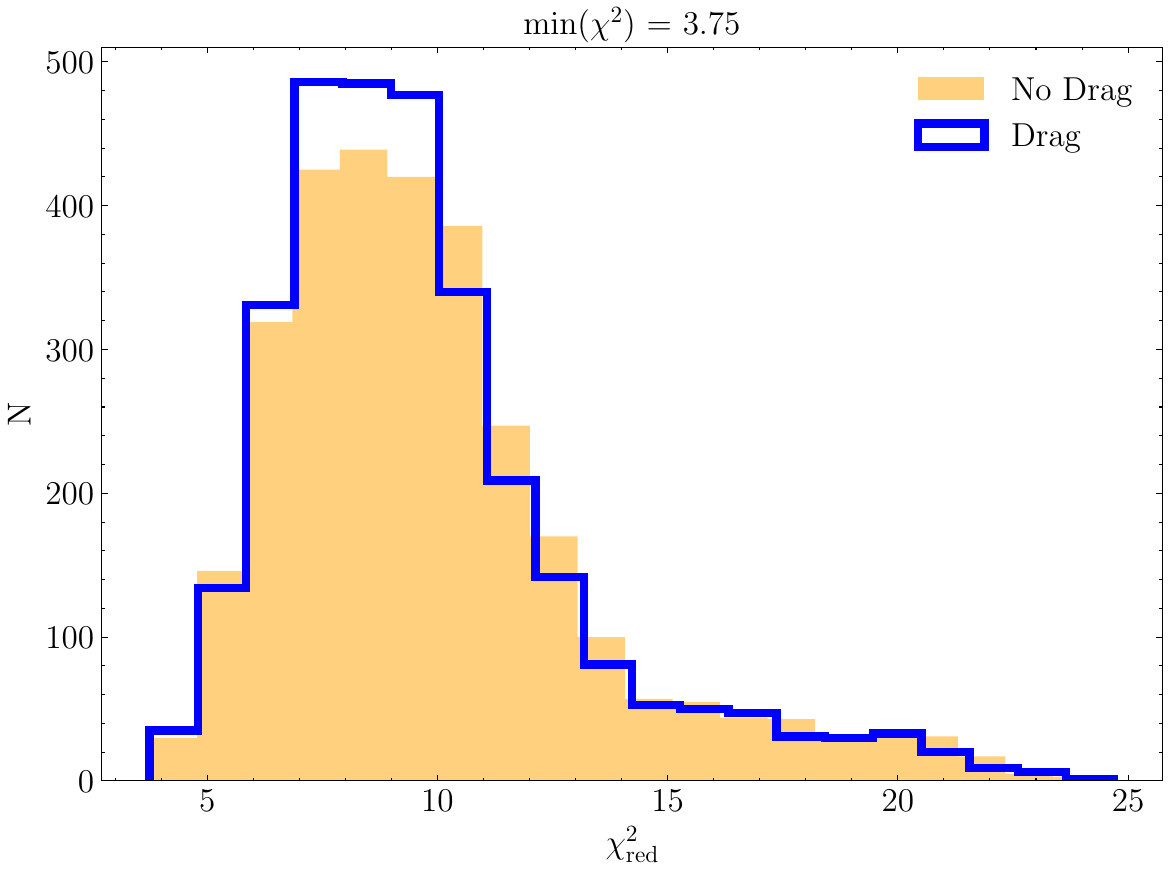}
            \caption{
            Histogram of the $\chi^2_{\text{red}}$ distribution of the 3000 realizations of the simulated cloud with initial conditions obtained from sampling the posterior distribution of the orbital fit of X7. 
            The solid blue line and orange area represent the results of the fiducial case and the one considering the drag force of the ambient medium, respectively. 
            The minimum value is $\chi^2_{\text{red}}=3.75$ in both cases.
            }
            \label{fig:Chi}
        \end{figure}

    \subsection{Comparison with observational data}

        Figure \ref{fig:angled_combi} shows a comparison of the morphological evolution of X7 observed in the L$'$ band and the simulated cloud during the period 2002-2021. 
        Each panel shows an image of the sky of an area of $1\arcsec\times~1\arcsec$, where Sgr~A* is located in the top left corner.
        In 2002, X7 had a much rounder shape than the simulated cloud, which is already elongated. 
        At later times, the observations of X7 display a more elongated morphology that visually tends to align with the simulated cloud. 
        This is clearer after 2015 when most of the observed shape of X7 is contained within the sky-projected simulated cloud.
        The agreement is best in the last epoch although this is not surprising since the initial velocity was selected so that tip of the cloud reproduces the position of X7 at this time. 

        \begin{figure}
            \centering
            \includegraphics[width=0.85\linewidth]{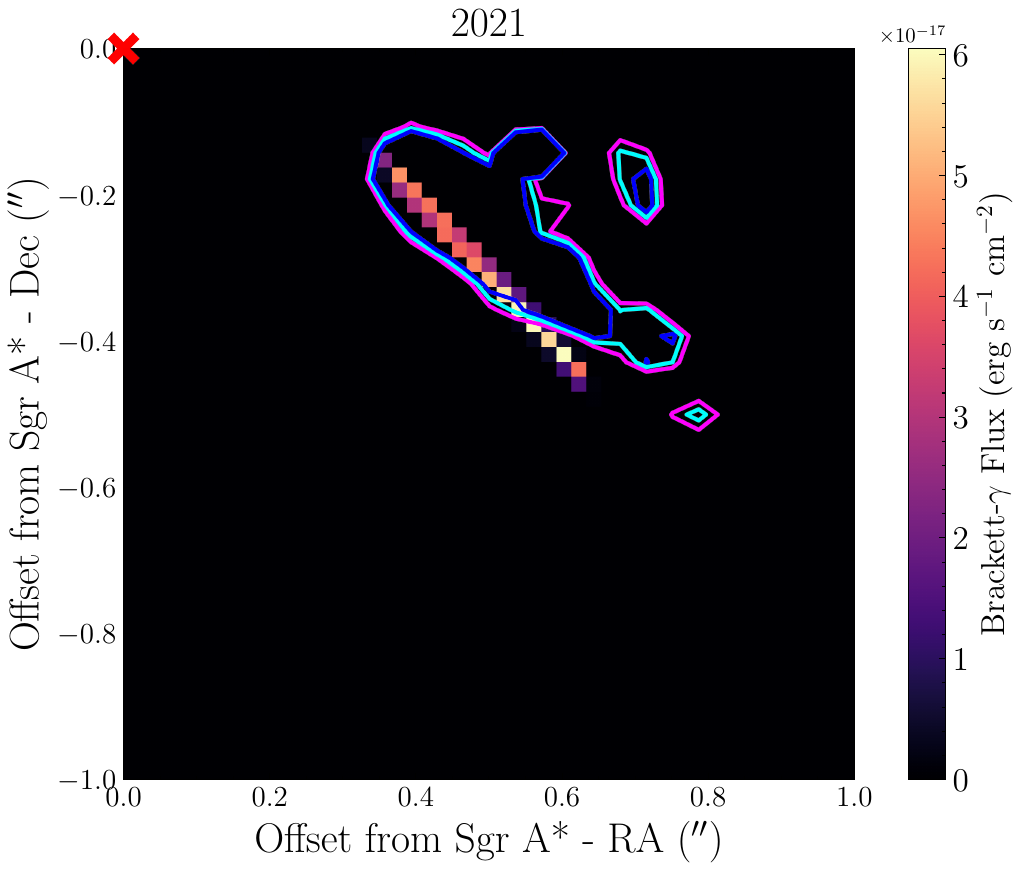}
            \caption{
            Comparison of the morphology of X7 and the simulated cloud through the observed (contours) and simulated (color map) Br$\gamma$ images in 2021. 
            The contour levels represent 0.25~(blue), 0.275~(cyan), and 0.30~(magenta) of the Br$\gamma$ emission maximum.
            The sky area shown is 1$\arcsec\times$~1$\arcsec$, with Sgr A* located in the top-left corner of the plot.
            }
            \label{fig:brgamma}
        \end{figure}

        To make a quantitative comparison, we calculated the reduced chi-square parameter ($\chi^2_{\text{red}}$) for the tip of the simulated cloud and X7 considering the three-dimensional position and velocity across every single observation epoch. 
        Additionally, we took into account the $1\sigma$ uncertainties in the orbital fit of X7. 
        Thus, we conducted 3000 realizations of this procedure sampling the initial conditions from the posterior distribution of the orbital fit of X7. 
        Figure \ref{fig:Chi} shows the $\chi^2_{\text{red}}$ distribution of this procedure. 
        This analysis also includes the impact of a hypothetical drag force due to the interaction of the cloud with the ambient medium (see Sect.~\ref{sec:drag}). 
        The distribution shows that the median value of $\chi^2_{\text{red}}\sim8$ and 
        a minimum value of $\chi^2_{\text{red}}=3.75$. 
        Overall, the model can reproduce the observed orbit of X7 to a large extent. 
        This result is not significantly affected by the drag force of the medium.
       
       To make a more appropriate comparison of the model with observational data, we synthesized the Br$\gamma$ emission from the simulated cloud. 
       Following Case B recombination theory, the Br$\gamma$ emissivity can be estimated following \cite{Schartmann_2015} as
        \begin{equation}
            j_{Br\gamma} = 3.44 \times 10^{-27} \left( \frac{T}{10^4 \text{ K}} \right)^{-1.09} \text{erg s}^{-1}\text{ cm}^{-3}.
        \end{equation}
        We assumed a temperature of $10^4$~K for the simulated cloud, which is typical for gas structures in the central parsec due to the strong ultraviolet radiation from the young, massive stars \citep{Zhao_2009, Gillessen_2012}.
        The total emissivity was calculated by multiplying by $n_\text{e}n_\text{p}$, where $n_\text{e}$ and $n_\text{p}$ are the electron and proton number density of the cloud. 
        We obtained the corresponding flux observed from Earth by integrating over the volume element and then scaling it by the inverse square law as
        \begin{equation}
            F_\lambda = \frac{\int{j_{\text{Br}\gamma}\ n_\text{e}n_\text{p}}\  dV}{4\pi R_0^2}
        ,\end{equation}
        where $V$ is the volume of the cloud, and $R_0 = 8300$~pc is the distance to the GC.
        Figure \ref{fig:brgamma} shows a comparison of the sky-projected observed and simulated Br$\gamma$ flux as seen from Earth in 2021. 
        Notice that most of the simulated emission is spatially contained within the observational contours of 0.3 of the maximum emission. 
        The simulated cloud is noticeably thinner, which could be a result of not taking into account the instrument point-spread function. 
        For completeness, a complete comparison of the evolution of the Br$\gamma$ flux during the period 2006-2021 is shown in Fig. \ref{brgammacombined}. 
        Quantitatively, the total flux obtained from our simulated cloud is $\sim1.28\times 10^{-15}~\text{erg~s}^{-1}~\text{cm}^{-2}$, whereas the total observed flux from X7 is $3.55 \times 10^{-15}~\text{erg~s}^{-1}~\text{cm}^{-2}$ in 2021 \citep{ciurlo2023}. 
        In summary, the model produces a cloud of similar morphology and emission of the same order of magnitude of the observations.

    \subsection{Secondary effects}
    \label{secondary}

        \begin{figure*}
            \centering
            \includegraphics[width=0.9\linewidth]{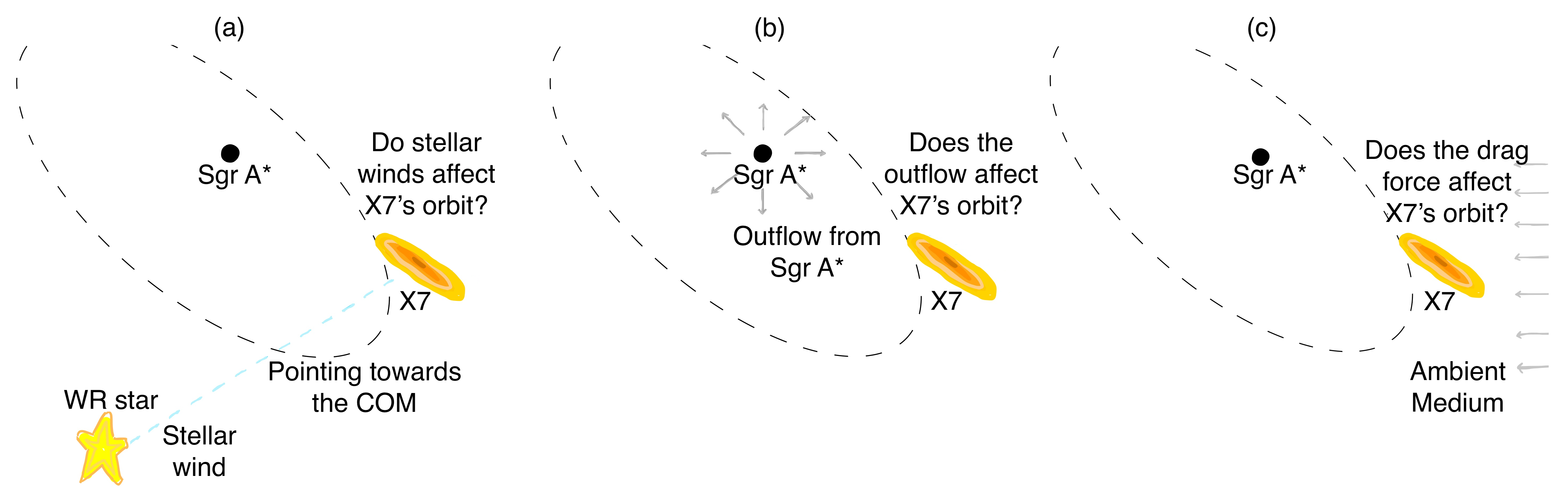}
            \caption{Schematic representation of the secondary effects that could affect the dynamical evolution of X7. (a) Ram pressure due to stellar winds pointing toward the center of mass of X7. (b) Ram pressure due to a spherical outflow from Sgr A*. (c) Drag force of the medium (the interstellar medium) in the vicinity of X7. Not to scale.}
            \label{fig:Schematic2}
        \end{figure*}

        \begin{figure}
            \centering
            \includegraphics[width=0.85\linewidth]{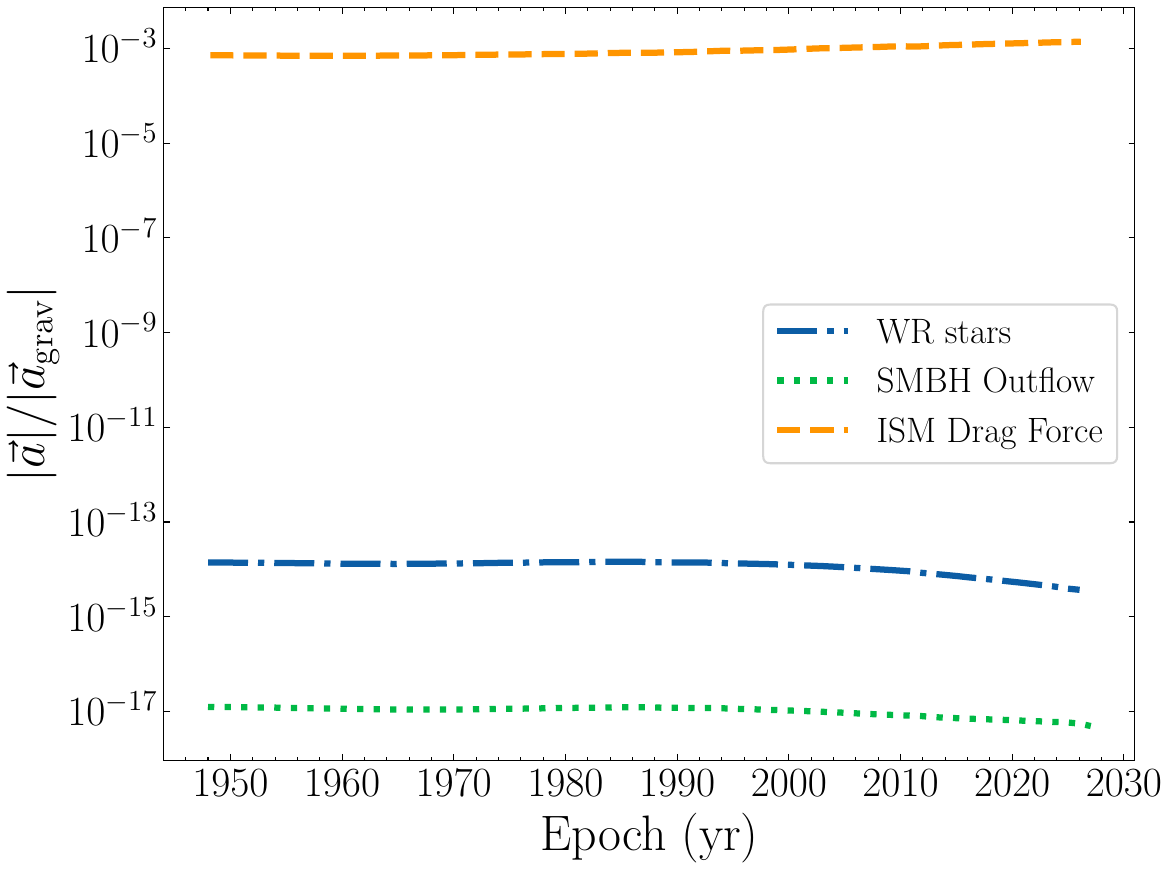}
            \caption{Ratio of magnitude of the acceleration on the simulated cloud due to secondary effects and gravity as a function of time.
            The effects of the stellar winds of the five closest WR stars, an outflow from Sgr A*, and the drag of the medium are shown as dashed orange, dotted green, and dotted-dashed blue lines, respectively.}
            \label{fig:secondary_effects}
        \end{figure}

        Up to now, our models have considered solely the effect of the gravitational field of Sgr~A*. 
        In this section we study the effects of other mechanisms that could play a role in the dynamical evolution of X7. 
        First, we estimated the impact of stellar winds from the WR stars. 
        Within the central parsec, there are $\sim$30 WR stars with strong mass loss of  $\sim$10$^{-5}$~M$_{\odot}$~yr$^{-1}$ launched at 500-2500~km~s$^{-1}$ \citep{Martins2007, cuadra2008}.
        Secondly, we investigated the effect of an outflow launched from the central region. 
        The accretion flow of Sgr~A* has been described successfully as a radiatively inefficient accretion flow due to both its inability to radiate its energy and the small fraction of the gas that ends up being accreted \citep[e.g.,][]{genzel_2010}. 
        In this scenario, the flow gets overheated and part of it can be launched as an energetic mass outflow \citep{blanford1999,mitchell2012}. 
        Assuming such an outflow indeed took place in the GC, we can also estimate its impact on the evolution of X7. 
        Lastly, we considered the effect of a drag force due to the interaction between the object and the ambient medium. 
        This can alter the orbits of the test particles that represent X7 in our model.
        Schematic representation of these effects are shown in Fig.~\ref{fig:Schematic2}. 

        \subsubsection{Stellar winds from Wolf-Rayet stars}
        \label{StellarWind}
            To date, $\sim$18\% of the stars in the central parsec of the GC have complete orbital solutions \citep{vonFellenberg_2022}. 
            For the rest, we used the solutions obtained by minimizing their orbital eccentricity (see Sect. \ref{MINECC}). 
            Then we found the closest WR stars to X7 since the year 1947 until 2021. 
            We considered only the five WR stars with closest approach to X7 in that period, namely E40, E48, E80, E81, and E83 \citep[in UCLA nomenclature: S3-5, IRS13 E4, S9-9, S9-283, and S10-5, respectively;][] {paumard2006}. 
            These stars have winds with mass-loss rates on the order of a few times $10^{-5}~\msunyr$ and velocities in the range 800-1000~$\kms$ \citep{Martins2007,cuadra2008}. 
            The ram pressure exerted by the stellar winds on X7 is given by $P_\text{w} = \rho_\text{w} v_\text{w,X7}^2$, where $v_\text{w,X7}$ is the net velocity of the wind acting on X7 taking into account the relative velocity of X7 and the WR star. The density of the stellar wind, assumed to be spherically symmetric and stationary, is denoted by $\rho_w$ and is given by
            \begin{equation}
                \rho_\text{w} = \frac{\dot{M}}{4\pi V_\text{w} R^2}
            ,\end{equation}
            where $\dot{M}$ is the stellar wind mass-loss rate, $V_\text{w}$ is the terminal velocity of the stellar wind, and $R$ is the distance from the star. 
            By adding an additional term as a source of acceleration due to the ram pressure in our simulation, we can estimate whether stellar winds from WR stars can affect the evolution of the cloud. 
            Then, the new equation of motion is given by
            \begin{equation}
                \frac{d^2\vec{r}}{dt^2} = -\frac{GM_\text{BH}}{r^2}\vec{\hat{r}} - \frac{\sigma_\text{c}}{m_\text{c}}\rho_\text{w} \left|\frac{dr}{dt}\right|^2 \vec{\hat{v}}
            ,\end{equation}
            where $\sigma_c$ and $m_c$ are the cross section in the direction of the net velocity vector $v_\text{w,X7}$ and mass of X7, respectively.
            Then, we simulated a simple model where the net stellar wind is pointed toward the center of mass of the cloud. 
            The results show that the motion of the simulated cloud does not change significantly. 
            Figure~\ref{fig:secondary_effects} shows the ratio of the stellar wind and the gravitational accelerations as a function of the simulation time. 
            Here it is possible to observe that this value remains roughly constant around $10^{-14}$. 
            This shows that their effect is negligible during this time period.

        \subsubsection{Outflow from the SMBH}
        
            We performed a similar analysis as in Sect. \ref{StellarWind} but considering a spherically symmetric outflow launched from the location of Sgr A*. 
            The exact mass-loss rate and velocity of the outflow were set to $\dot{M}_\text{BH} = 10^{-4}~\msunyr$ and $v_\text{out} = 10^9$~cm~s$^{-1}$, respectively and motivated by numerical simulations of this process \citep[e.g.,][]{Cuadra_2015}. 
            Our simulation shows that the evolution of the orbit of the simulated cloud is not affected significantly by the outflow. 
            The relative magnitude of this acceleration compared to the gravitational acceleration as a function of time is shown in Fig.~\ref{fig:secondary_effects}. 
            The value is on the order of 10$^{-17}$, which is even smaller than the effect of the stellar winds and, therefore, is also negligible. 
            Moreover, even a much stronger outflow will have a negligible effect on the orbit of X7. 
            It is relevant to mention that this result is in agreement with the analysis in \cite{ciurlo2023}, as they argued that the morphology of X7 cannot be explained by an outflow from Sgr~A*.

        \subsubsection{The effect of a static drag force}
        \label{sec:drag}
        
            To account for the relative motion of the ambient medium we also investigated the effect of a static drag force acting against the motion of the cloud. 
            This has been taken into account previously for the G2 object \citep{Madigan_2017,Calderon_2018}, and later observed by \citep{Gillessen_2019}. 
            In principle, this would cause deviations from the Keplerian orbits of the test particles of the simulated cloud. 
            In this case, the equation of motion becomes 
            \begin{equation}
                \frac{d^2 \vec{r}}{dt^2} = -\frac{GM_\text{BH}}{r^2}\vec{\hat{r}} - \frac{\sigma_\text{c}}{m_\text{c}} \rho_\text{ISM}(r) \bigg|\frac{d\vec{r}}{dt}\bigg|^2 \vec{\hat{v}}
            ,\end{equation}
            where $G$ is the gravitational constant, $M_\text{BH}$ is the mass of Sgr A*, $m_\text{c}$ and $\sigma_c$ are the mass and cross section of the cloud, and $\vec{\hat{r}}$ and $\vec{\hat{v}}$ are the unit vectors in the radial and velocity directions, respectively. 
            The density of the ambient medium ($\rho_\text{ISM}$) is based on the model from \cite{Yuan_2003}, which reproduces the \textit{Chandra} X-ray observations and is consistent with the latest model by \cite{2017_Roberts}. 
            This density profile is given by
            \begin{equation}
                \rho_\text{ISM}(r) = 10^{-22} \left(\frac{1.7 \times 10^{17}}{r}\right)^\alpha~\text{g}~\text{cm}^{-3}
            ,\end{equation} 
            where $\alpha$ is the power-law index set to $\alpha=1$ based on the analysis of \cite{burkert_2012}, although we also tested $\alpha = 0.5$ and $\alpha = 1.5$ and found no substantive change in the result. 
            Moreover, the mass of the cloud is constrained to $\sim$50~$\mearth$ \citep{ciurlo2023}.
            Then, assuming that the masses of the particles are identical, the mass of each particle was determined. 

            In this case, the drag force affects the orbit of the test particles. 
            Figure \ref{fig:secondary_effects} shows the ratio of the drag force and gravitational accelerations as a function of time. 
            Here it can be seen that the value is $\sim$10$^{-3}$, which is much larger than in the previous cases. 
            This effect results in changes in the position of the tip of the simulated cloud that increases the three-dimensional separation of the simulated tip and X7 in 2021. 
            Specifically, the fiducial model results in a separation of $\sim$50~au while the calculations with the drag force shows it to be $\sim$130~au. 
            However, there is no significant change in the length of the cloud. 
            Although $\Delta r_{\text{tip}}$ increases, the result is still within the margin of the errors. 
            Thus, this effect does not impact significantly the overall results. 
            It is to be noted that a non-static drag force could have slightly different effects \citep{Madigan_2017}.

\section{Discussion}
    \label{discuss}

        We have shown that ejecta from S33/S0-30 provided with an ejection velocity of $\sim$610~$\kms$ in 1947 can reproduce the orbit of X7 such that the tip of the simulated cloud and the observed tip of X7 in 2021 have a separation ($\Delta r_{\text{tip}}$) of around 50 au. 
        We have also shown that an initially elongated cloud with an initial velocity gradient can reproduce the observed orientation of X7 (see Fig.~\ref{fig:ellipse}). 
        In this section we interpret these results based on our proposed hypotheses for the origin of X7.

        \subsection{A star in an extensive mass-losing phase}
            
            We started with the hypothesis that X7 formed when a star in the past 200 years went through a phase of excessive mass loss such as a LBV phase. 
            The abundance of O-type and WR stars in the GC support this idea, since it is possible for a O-type star to go through a LBV phase before becoming a WR star \citep[e.g.,][]{2007_Crowther}. 
            The WR stars in the GC have a mass-loss rate of a few times $10^{-5}~\msunyr$ \citep{Martins2007}. 
            With this rate, it would take approximately 15 years to lose an amount of mass comparable to the mass of X7. 
            However, the best candidate to be the origin of X7, S33/S0-30, is a B-type star whose mass-loss rates are not as high \citep[$\lesssim$10$^{-8}~\msunyr$;][]{Martins_2008}. 
            Even if the mass-loss rates were to go an order or two higher, the timescale needed would be too long to generate a single coherent structure comparable to X7. 
            Moreover, such a mass loss will occur in a spherically symmetric manner, whereas our model favors an ellipsoidal ejecta. 
            Based on this, it is very unlikely that X7 originated in this way.

            \begin{figure}[h]
                \centering
                \includegraphics[width=0.95\linewidth]{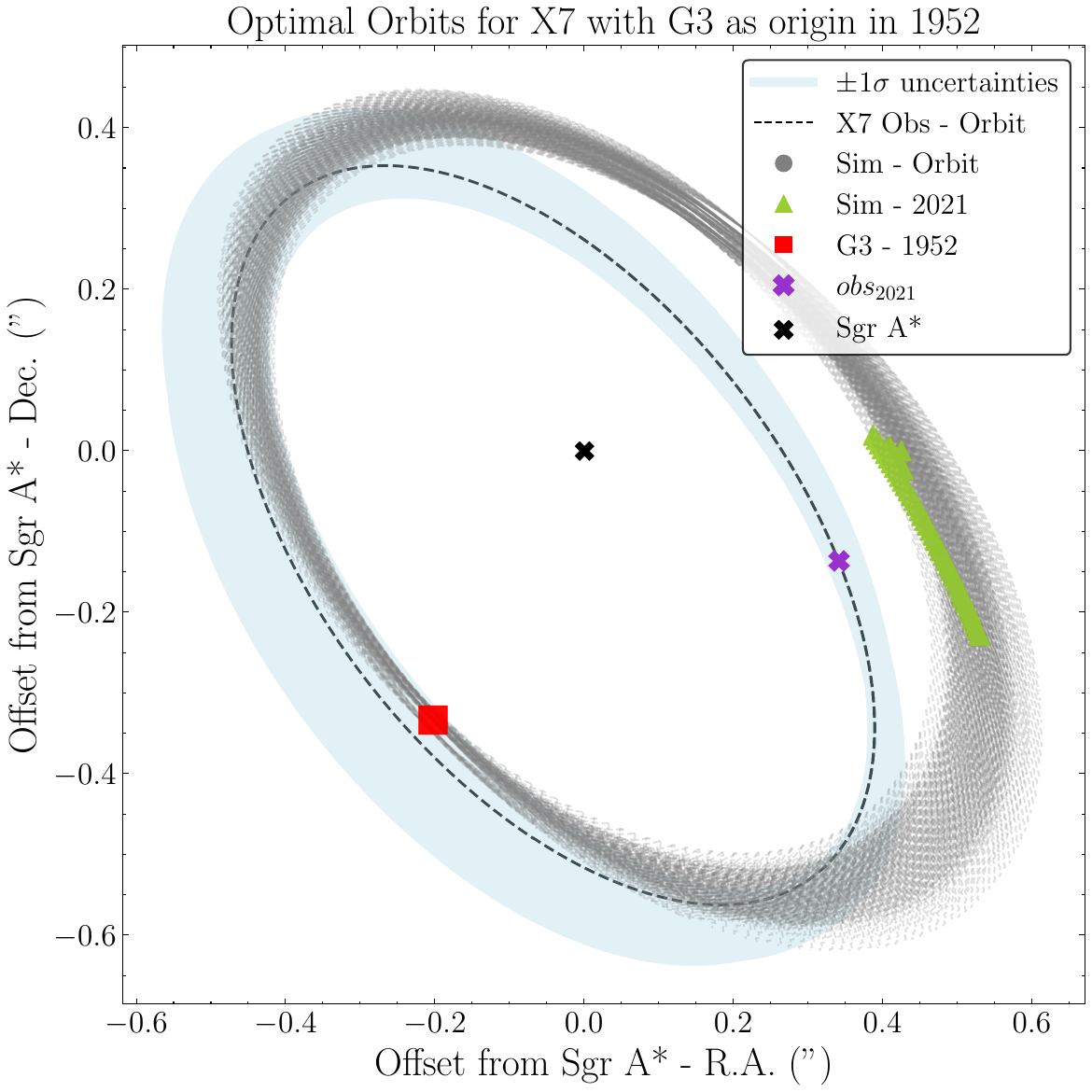}
                \caption{
                Comparison of the sky-projected orbits and position in 2021 of X7 and the tip of the simulated clouds from G3 in 1952. 
                The dashed black line shows the best-fit orbit of the tip of X7, the light blue shaded region shows the uncertainties with a 68$\%$ confidence interval, and the purple cross marks its position in 2021 \citep{ciurlo2023}. 
                The dotted gray lines show the orbit of the simulated clouds for each of the optimal velocities. 
                The red squares and light green triangles show the position of the tip of the simulated clouds in 1952 and 2021, respectively.
                }
                \label{fig:X7_from_G3}
            \end{figure}
        
    \subsection{Ejecta from a stellar merger: Are X7 and G3 dynamically linked?}
    \label{G3_X7}

        Since X7 and G3 have similar orbits and emission characteristics they have been hypothesized to have been formed in an EKL-induced merger scenario where G3 is the merger product and X7 is the ejected mass \citep{ciurlo2023}. 
        Notice that even the orientations of the angular momentum of X7 and G3 are strikingly similar (see~Fig. \ref{fig:Angular Momentum}). 
        To test this hypothesis, we conducted an analysis similar to that done for S33/S0-30 and X7. 
        We extrapolated the orbit of G3 and X7 backward in time in order to find their closest separation over the last 200~yr period. 
        As a result, we find that this occurred on two occasions: in $\sim$1880 and $\sim$1952. 
        However, we notice that G3 is always orbiting ahead of X7 on the sky. 
        Their three-dimensional separation in the closest encounters is larger than 1200~au. 
        Even taking the cases of the closest encounters as the origin of the simulated cloud, we ended up obtaining significant differences between the expected position of the tip of the simulated cloud and X7 in 2021. 
        In the cases of 1880~yr and 1952~yr, we find that the separations are $\sim$40000~au and $\sim$1200~au between their tips, respectively. 
        When we consider the uncertainties in the orbit of X7, the three-dimensional separation remains of the same order of magnitude in both cases. 
        Moreover, the sky-projected position for the best cases in 1952 are offset beyond the uncertainties as shown in Fig. \ref{fig:X7_from_G3}.

        These results show that ejecta from G3 in 1880 or 1952 cannot be placed on an orbit like the one observed for X7. 
        Thus, it is difficult to reconcile the idea of X7 being ejecta from G3 over the last 200~yr. 
        We note that this conclusion relies on the current orbital estimates of X7 and G3, which have relatively large uncertainties due to short orbital phase coverage \citep[$\lesssim$10\%;][]{ciurlo2023}. 

    \subsection{Grazing collision of a star with a field object}

        The case of the grazing collision of a star and a field object such as a stellar-mass black hole or even a Jupiter-mass object has been discussed in detail by \cite{ciurlo2023}. 
        The collisions between red giants and compact remnants have also been invoked to explain the depletion of red giants within the inner 10\arcsec~of the GC \citep{DaviesMB2011}. 
        In principle, a grazing collision of a red giant with any such field object could account for the mass of X7 by stripping enough material from the star's atmosphere \citep{ciurlo2023}. 
        This is an interesting prospect since our simulations show that an initially elongated ejecta from S33/S0-30 with an initial velocity of 610 $\kms$ can indeed be placed on an orbit similar to X7's, as well as reproduce its observed tail orientation (see Fig. \ref{fig:ellipse}). 
        However, if that were the case, the star would be left in an agitated state and would only settle down on a Kelvin-Helmholtz timescale \citep{ciurlo2023}. 
        To date, there is no account of such effects for S33/S0-30, but future observations might be able to detect this.
        
        To assess how feasible this scenario is, we need to ask how often (grazing) collisions happen in the GC. 
        There have been studies dedicated to estimating the stellar collision rate, motivated by the missing red giant problem \citep{2020_Rose, 2022_Rose, 2023_Rose, Amaro_Seoane_2023}. 
        These studies estimated this rate to be about $10^{-4}$~yr$^{-1}$ in a volume of 100~Mpc for a Milky Way like nucleus. 
        \cite{Balberg_2023} developed a probabilistic approach to study the rates of collisions in a stellar cluster with a SMBH at its center. 
        This work predicted collision rates in the range 10$^{-6}$-$10^{-5}$~yr$^{-1}$, smaller, as they referred to destructive, i.e., physical collisions as opposed to just stripping the outer layers of the stars. 
        However, none of these works considered the potential collisions between a field object such as a stellar-mass black hole and a star. 
        The central parsec is thought to host a population of compact objects, including stellar-mass black holes \citep{1993_Morris,mouawad2005,zhao2022}.
        Recent work by \cite{Haas_2025} calculated the radial density profile of the black hole cluster in the GC based on observations of the stellar populations. 
        Their estimations suggest that at a distance of 0.015~pc from Sgr~A*, which is approximately the location of X7, a star of 10~M$_{\odot}$ should have gone through a collision with a black hole on a timescale of 55~Myr, while a more massive star of 50~M$_{\odot}$ would do so on a shorter timescale of 5~Myr. 
        Moreover, it is indeed feasible that such collisions have already taken place for some of the stars that inhabit the region, as the young population age is on the order of 2.5-5.8~Myr \citep{lu2013}. 
        Under this scenario, let us consider the star S33/S0-30 and assume its mass to be 10~M$_{\odot}$. 
        Currently, it is possible to observe $>$100 B-type stars like it in the region. 
        Then, we would expect one collision between a B-type star and a stellar-mass black hole every 0.05~Myr. 
        If we consider that a stripped filament, such as X7, lives for 100~yr, the chances of detecting it would be 0.2\%. 
        Notice that this value should be interpreted as a lower limit since S33/S0-30 could have been more massive in the past and would have lost mass due to this kind of interaction.
        Furthermore, we note that the calculations by \cite{Haas_2025} do not consider a preferential geometry. 
        In the case of the young stars in the GC, the collision timescale could be shorter if we weigh in the fact that a significant fraction of the stars have been formed on disk structures \citep{paumard2006, Gillessen_2017, vonFellenberg_2022}. 
        Moreover, it is to be noted that the grazing collision timescales should be even shorter than those computed by \cite{Haas_2025} for more direct collisions, given the larger cross section for grazing collisions. 
        Thus, although uncertain, the collision rate estimations between a field object and a star are compatible with X7 having formed in one such event.

    \subsection{An infalling gas filament}

        \begin{figure}
            \centering
            \includegraphics[width=0.85\linewidth]{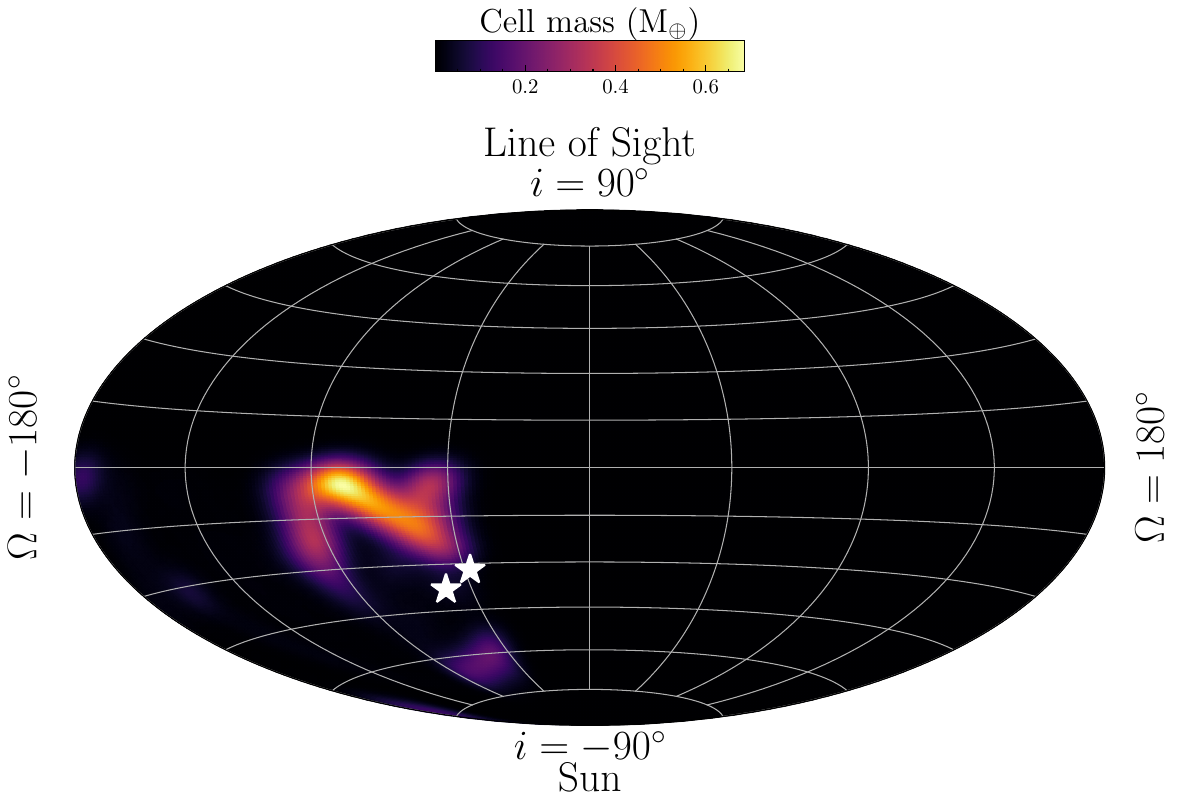}
            \caption{Density map of the gas structures in the \texttt{RAMSES} simulation for the stellar winds from the WR stars feeding the black hole \citep{calderon2025}. The gas cells with $T < 10^5$~K, inclination $-90^\circ < i < 0^\circ$, and longitude of ascending node $-180^\circ < \Omega < 0^\circ$ are considered to find coherent structures with a similar orientation of the orbital angular momentum to that of X7 and G3. The color bar refers to cell mass. None of the gas cells form any coherent clumps that could evolve into X7.}
            \label{fig:yt_ramses}
        \end{figure}

        An additional hypothesis not studied yet is considering X7 as a gaseous clump or stream resulting from the many stellar wind collisions that take place in the vicinity of Sgr~A*. 
        Although single stellar wind collisions have been shown to produce only light clumps \cite[$\lesssim$0.01~M$_{\oplus}$;][]{calderon2020b}, in principle the simultaneous interaction of more stellar winds could result in larger and/or more massive structures. 
        Hydrodynamic simulations of the WR stellar winds feeding Sgr~A* have been performed by many authors \citep{cuadra2005,cuadra2006,cuadra2008,Cuadra_2015,russell2017,ressler2018,Calderon_2020,ressler2020,solanki2023,balakrishnan2024,calderon2025}.
        Here, we analyzed the hydrodynamic models developed by \cite{calderon2025} in order to search for cold and gaseous structures that might share properties with X7.
        We selected an output of the simulation that corresponds to the last observation of X7, i.e., $t=2021$. 
        Then, we analyzed the region within 10\arcsec~($\sim$0.4~pc) from Sgr A*. 
        We imposed constraints on gas temperature of $<$10$^5$~K, inclination $-90^\circ < i < 0^\circ$, and longitude of ascending node $-180^\circ < \Omega < 0^\circ$ in order to find cold structures oriented similarly to that of X7. 
        The results of this analysis are shown as a Hammer projection in Fig.~\ref{fig:yt_ramses}. 
        Here it is possible to observe that there seem to be cells with similar angular momentum direction to X7's. 
        However, we checked whether such cells form a coherent spatial structure and found that these are not spatially connected. 
        Thus, it is unlikely that X7 has originated from WR stellar wind interactions.
        
\section{Conclusion}
\label{conclusion}
    We have studied the dynamical evolution of the source X7 in order to constrain its origin. 
    Under the assumption that this object is purely a gas and dust feature, we tested three scenarios for its formation: from the wind of a star going through a major mass-loss episode, as the ejecta of a binary merger, or as the ejecta from the grazing collision of a star with a stellar-mass black hole or Jupiter-mass object. 
    To test these hypotheses, we analyzed 195 stars with observationally constrained orbits within the central parsec of the GC. 
    We searched for suitable progenitors with a closest approach to X7 of $<$1000~au, relative velocities of $<$1000~$\kms$ over the last 200 years, and a similar angular momentum orientation. 
    With these criteria, we found only one suitable candidate: the star S33/S0-30 \citep{2009_gillessen, Gillessen_2017, vonFellenberg_2022}. 
    This star and X7 had their closest approach in the year 1947 with a minimum separation and relative velocities of $\sim$600~au and $\sim$500 $\kms$, respectively. 
    We modeled the evolution of hypothetical ejecta with a set of test particles launched from the position of S33/S0-30 in 1947 with an initial velocity that minimizes the difference between the three-dimensional distance and the relative velocity between the simulated cloud and X7. 
    This analysis shows that:

    \begin{enumerate}
        \item It is unlikely that X7 formed from a stellar wind. 
        No star with sufficiently high mass loss has been close enough to its position over the last 200 years.
        \item Despite their similar orbits, X7 and G3 do not seem to be related: extrapolating their positions into the past does not result in close encounters, and ejecta from G3 cannot form the orbit of X7. 
        It is relevant to bear in mind that this conclusion relies on the age of X7 being less than its orbital period of $\sim$200 years.
        \item Initially elongated ejecta of length $L=200$ au from the star S33/S0-30 with an initial velocity of 610~$\kms$ and an initial velocity gradient from tip to tail can be placed on a similar orbit to X7's, and can reproduce the current position of its tip. 
        Additionally, the length of such ejecta is $\sim$3350~au in 2021, which is consistent with the observational data.
        \item Secondary effects such as stellar wind ram pressure and a hypothetical feedback event from the  SMBH are not significant enough to affect the evolution of X7. 
        A static interstellar medium drag can indeed change the orbit of X7, but it does not affect its dynamics significantly over the simulated timescale of $\sim$80~yr.    
    \end{enumerate}
        
    \noindent From these results, we speculate that a grazing collision between S33/S0-30 and a field object such as a stellar-mass black hole or a Jupiter-mass object could have produced the ejecta that corresponds to X7. 
    Furthermore, we have argued that the current constraints on the presence of a population of stellar-mass black holes in the vicinity of Sgr~A* make grazing collisions a plausible occurrence that could account for the production of X7.
    
    It will be interesting to observe the star S33/S0-30 for any extended dusty envelope, which would further strengthen our results. 
    As X7 approaches the pericenter passage, it will elongate more and begin fragmenting, which has already been observed. 
    Our simulations are fairly accurate before the pericenter passage. 
    To reveal more about the tidal interaction and post-pericenter evolution, numerical hydrodynamic simulations are required. 
    Future observations will reveal more interesting properties with better constraints.
    
\begin{acknowledgement}
    We would like to thank the anonymous referee for the comments that helped to improve this article.
    WS and JC acknowledge the kind hospitality of the Max Planck Institute for Astrophysics where part of this work was developed. 
    The research of WS, DC, and SR has been funded by the Deutsche Forschungsgemeinschaft (DFG, German Research Foundation) under Germany’s Excellence Strategy - EXC 2121 - ``Quantum Universe” - 390833306. 
    Since 01.10.2024, DC has been funded by the Alexander von Humboldt Foundation.
    SR has also been supported by the Swedish Research Council (VR) under grant number 2020-05044, by the research environment grant ``Gravitational Radiation and Electromagnetic Astrophysical Transients” (GREAT) funded by the Swedish Research Council (VR) under Dnr 2016-06012, by the Knut and Alice Wallenberg Foundation under grant Dnr. KAW 2019.0112, and by the European Research Council (ERC) Advanced Grant INSPIRATION under the European Union’s Horizon 2020 research and innovation programme (Grant agreement No. 101053985).  
    JC acknowledges financial support from ANID -- FONDECYT Regular 1211429 and 1251444, and Millennium Science Initiative Program NCN$2023\_002$.
    The analysis and visualizations in this work made use of Python libraries, Numpy \citep{2020_Harris} and Matplotlib \citep{2007_Hunter}. 
    \texttt{RAMSES} hydrodynamical simulation data analysis was carried out using the Python library YT \citep{YT_2011}. 
\end{acknowledgement}

\section*{Data availability}
Available on reasonable request to the corresponding author.

\bibliographystyle{aa}
\bibliography{X7}

\appendix

    \section{Spherical initial configuration}
    \label{spherical}

\begin{figure}[h]
            \centering
            \includegraphics[width=1\linewidth]{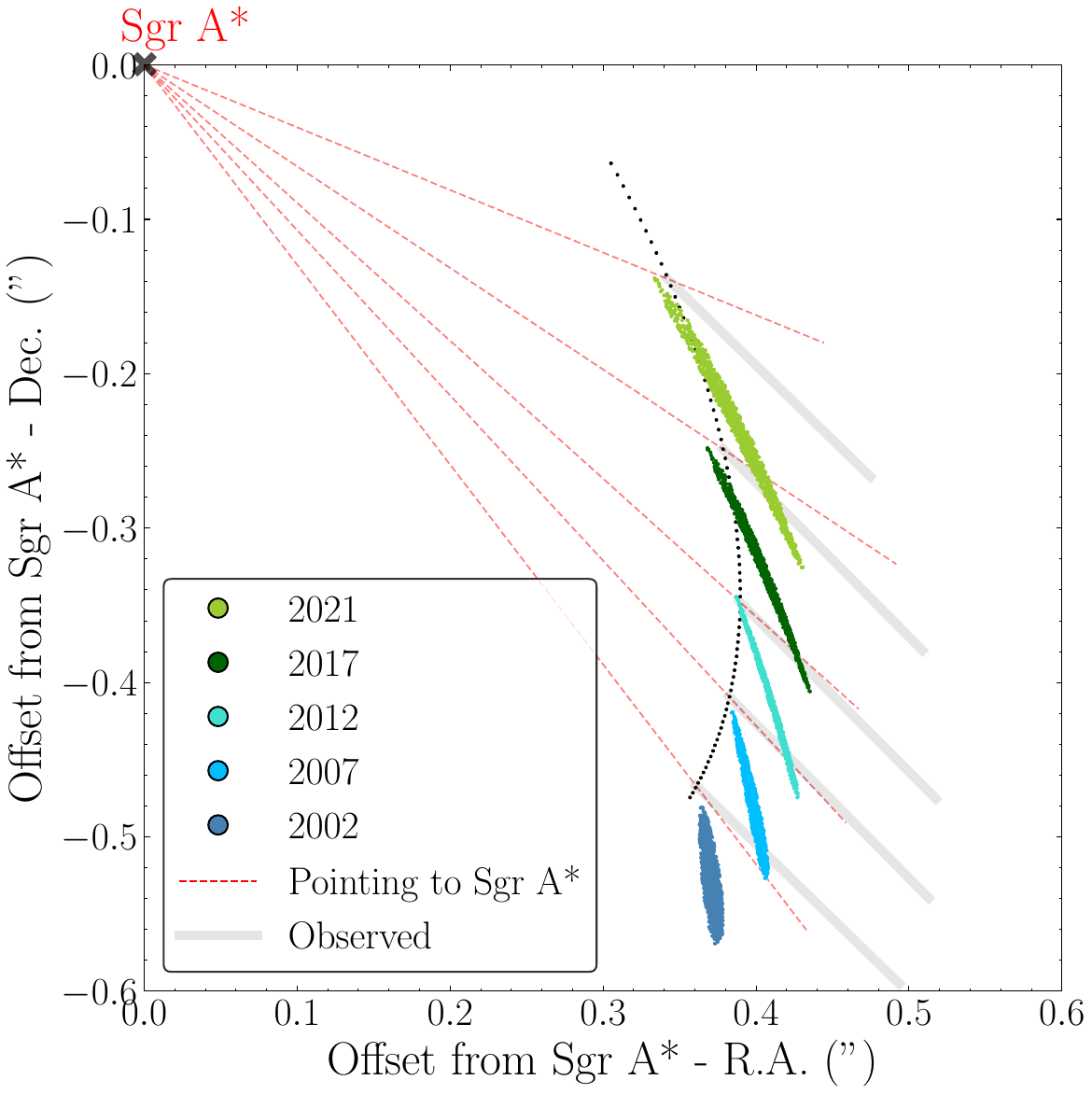}
            \caption{
            Comparison of the orientation of the initially spherical simulated cloud and X7. 
            The sky-projected morphology of the simulated cloud is shown at $t=2002, 2007, 2012, 2017$, and $2021$~yr as dark blue, light blue, cyan, dark green, and light green dots, respectively. 
            The gray shaded lines show the observed orientation of X7 \citep{ciurlo2023}.
            The dashed red line represents the lines connecting Sgr A* and the location of the tip of X7 at a given epoch. 
            Sgr~A* is located in the top-left corner.
            }
            \label{fig:orientation_tail}
        \end{figure}

        In the main text, we have presented the result of simulations that considered an initially ellipsoidal cloud as that showed the best agreement with the observations. 
        For completeness, here we include the results considering a uniform initially spherical cloud.
        This simulation resulted in a three-dimensional separation between the simulated cloud and the observed tips of X7 of $\Delta r_{\text{tip}}\sim 116$~au in 2021.
        The length of the simulated cloud was $\sim$2000~au in 2021 whereas the observations show a length of $\sim$3300~au \citep{ciurlo2023}. 
        However, the main discrepancy comes from the orientation of the tail of the simulated cloud and X7. 
        Figure \ref{fig:orientation_tail} shows the sky-projected morphology of the simulated cloud at $t=2002, 2007, 2012, 2017,$ and $2021$~yr and the orientation of the observed structure of X7 \citep{ciurlo2023}. 
        The disagreement between the orientation of the simulated and observed structures is clear. 
        The discrepancy is around $\sim 20^{\circ} - 40^{\circ}$ during the epoch 2002-2021~yr. 
            
        Moreover, we tested the initially spherical cloud case with a three-dimensional Gaussian distribution structure as well (see Table \ref{tab:SIMRUNS}). 
        This case resulted in $\Delta r_\text{tip}\sim$250~au, and relative velocities of $\sim$100~$\kms$ in 2021. 
        Additionally, the mismatch of the simulated cloud and X7 orientation was even more prominent.

    \section{A different orbit for X7}\label{peissker2024_orbit}

\begin{figure}[h]
            \centering
            \includegraphics[width=\linewidth]{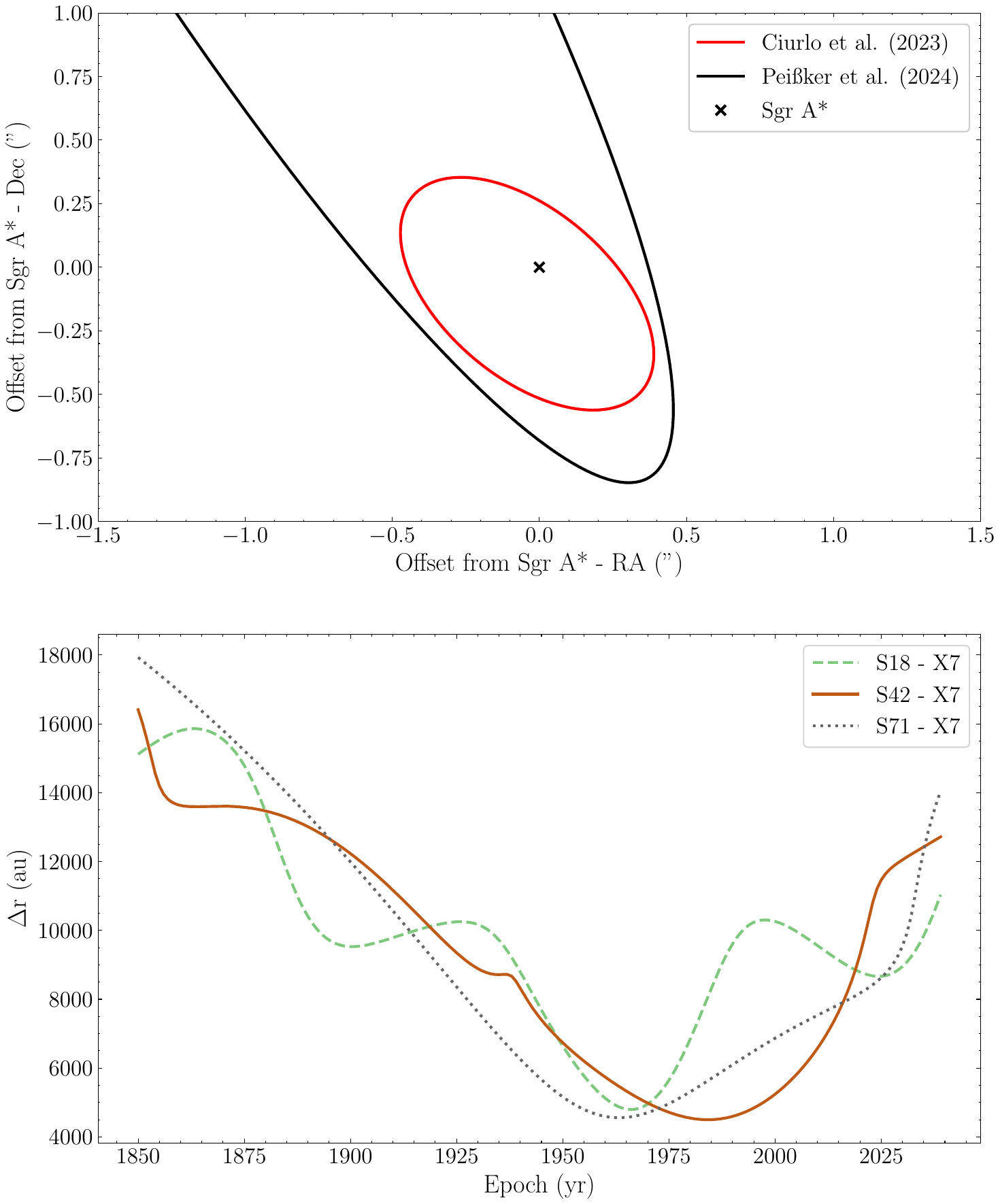}
            \caption{
            Top panel: Comparison of the best fitted orbits of X7 projected on the sky by \cite{ciurlo2023} and \cite{Peissker2024}, which are shown as solid black and red lines, respectively. 
            Bottom panel: Three-dimensional separation between X7 and candidate stars, used to constrain its origin based on an alternative orbital solution for X7 \citep{Peissker2024} as a function of time. 
            The three stars shown, S18, S42, and S71, have fully constrained orbits \citep{vonFellenberg_2022}.  
            Notice that their closest approach to X7 over the past 200 years is between 1950 and 2000 at a three-dimensional separation $>$4000 au.
            } 
            \label{fig:peissker2024}
        \end{figure}

        \cite{Peissker2024} derived a different orbit for X7 when compared to the one by \cite{ciurlo2023}. 
        A comparison between these orbits can be seen on the top panel of Fig. \ref{fig:peissker2024}. 
        It is clear that there is a large discrepancy between the orbits inferred by \cite{ciurlo2023} and \cite{Peissker2024}. 
        Although the uncertainties in \cite{Peissker2024} are smaller by an order of magnitude, the key reason for this different orbit is their L$^\prime$-band data point from 1999, which favors this newly derived orbit. 
        
        To test if the results of our work are affected by the X7 orbit choice, we repeated the analysis but making use of the orbit derived by \cite{Peissker2024}.
        The calculations show that the closest stars to X7 within the last 200 years are at least 4000~au apart (see the bottom panel of Fig. \ref{fig:peissker2024}). 
        Bear in mind that our original separation threshold was 1000~au. 
        As in this case the separation is at least four times larger, the use of a different orbit makes a relation to any of the star candidates less likely. 

\section{Simulated morphology evolution}

\begin{figure*}
\centering
\includegraphics[width=0.85\textwidth]{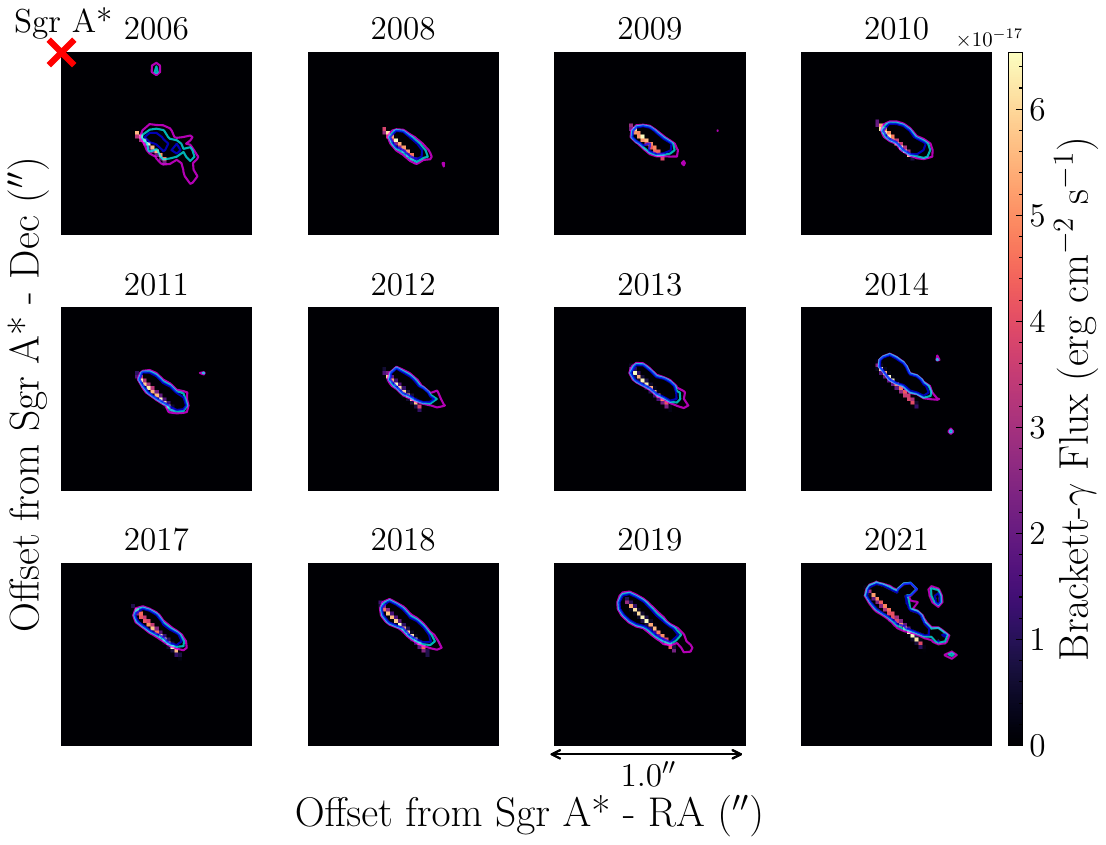}
    \caption{Comparison of the morphological evolution of X7 and the simulated cloud. 
    The panels show sky-projected 1\arcsec$\times$~1\arcsec Br$\gamma$ images and highlight the emission from X7 with contours in the period 2006-2021.
    The simulated cloud is overlaid with colored markers that encode the simulated Br$\gamma$ flux. 
    The position of Sgr A* is in the top-left corner of each panel. 
    North is up and east to the left.
    }
    \label{brgammacombined}
\end{figure*}

        For completeness, we analyzed all the Br$\gamma$ observations available and compared them with the simulated emission. 
        Figure~\ref{brgammacombined} shows the sky-projected Br$\gamma$ flux observed from Earth through the epochs $t=2006$, 2008, 2009, 2010, 2011, 2012, 2013, 2014, 2017, 2018, 2019, and 2021. 
        Notice that the observations contain most of the distribution of the simulated flux, and the emission level is of the same order as the reported values as it is shown in the main text.

\end{document}